\documentclass[english,structabstract]{aa}
\usepackage{mathptmx}
\usepackage[T1]{fontenc}
\usepackage[utf8]{inputenc}
\usepackage{color}
\usepackage{babel}
\usepackage{graphicx}
\usepackage[unicode=true]
 {hyperref}
\usepackage{breakurl}

\makeatletter
\bibpunct{(}{)}{;}{a}{}{,} % to follow the A&A style
\usepackage{hyperref}
\usepackage{longtable}
\usepackage{lscape}
\newcommand{\teff}{T_{\mathrm{eff}}}
\newcommand{\logg}{\log g}
\newcommand{\feh}{\left[\mathrm{Fe}/\mathrm{H}\right]}
\newcommand{\ssbot}{s_{\mathrm{bot}}}

\newcommand{\dss}{\Delta s}
\newcommand{\vzrms}{v_{z,\mathrm{rms}}}
\newcommand{\vzrmsp}{v_{z,\mathrm{rms}}^{\mathrm{peak}}}
\newcommand{\nab}{\vec{\nabla}}
\newcommand{\nsad}{\vec{\nabla}_{\mathrm{sad}}}

\newcommand{\hav}{\left\langle \mathrm{3D}\right\rangle}
\newcommand{\havz}{\hav_z}

\newcommand{\ttau}{T(\tau)}
\newcommand{\lmlt}{l}

\newcommand{\am}{\alpha_{m}}
\newcommand{\amlt}{\alpha_{\mathrm{MLT}}}

\titlerunning{The Stagger-grid -- III. Mixing length connection}
\authorrunning{Z. Magic et al.}
\usepackage{txfonts}

\makeatother

\begin{document}

\title{The \textsc{Stagger}-grid: A grid of 3D stellar atmosphere models}

\subtitle{III. The relation to mixing-length convection theory%
\thanks{Appendix is available in electronic form at \protect\href{http://www.aanda.org}{http://www.aanda.org}%
}%
\thanks{Full Table A.1 is available at the CDS via anonymous ftp to \protect\href{http://cdsarc.u-strasbg.fr}{cdsarc.u-strasbg.fr}
(\protect\href{http://130.79.128.5}{130.79.128.5}) or via \protect\href{http://cdsarc.u-strasbg.fr/viz-bin/qcat?J/A+A/???/A??}{http://cdsarc.u-strasbg.fr/viz-bin/qcat?J/A+A/???/A??},
as well as at \protect\href{http://www.stagger-stars.net}{www.stagger-stars.net}.%
}}

\author{Z. Magic\inst{1,2}, A. Weiss\inst{1} \and  M. Asplund\inst{2}}

\institute{Max-Planck-Institut für Astrophysik, Karl-Schwarzschild-Str. 1, 85741
Garching, Germany\\
\email{magic@mpa-garching.mpg.de} \and  Research School of Astronomy
\& Astrophysics, Cotter Road, Weston ACT 2611, Australia}

\offprints{magic@mpa-garching.mpg.de}

\date{Received ...; Accepted...}

\abstract{}{We investigate the relation between 1D atmosphere models that
rely on the mixing-length theory and models based on full 3D radiative
hydrodynamic (RHD) calculations to describe convection in the envelopes
of late-type stars.}{The adiabatic entropy value of the deep convection
zone, $\ssbot$, and the entropy jump, $\Delta s$, determined from
the 3D RHD models, were matched with the mixing-length parameter,
$\amlt,$ from 1D hydrostatic atmosphere models with identical microphysics
(opacities and equation-of-state). We also derived the mass mixing-length
parameter, $\am$, and the vertical correlation length of the vertical
velocity, $C\left[v_{z},v_{z}\right]$, directly from the 3D hydrodynamical
simulations of stellar subsurface convection.}{The calibrated mixing-length
parameter for the Sun is $\amlt^{\odot}\left(\ssbot\right)=1.98$.
For different stellar parameters, $\amlt$ varies systematically in
the range of $1.7-2.4$. In particular, $\amlt$ decreases towards
higher effective temperature, lower surface gravity and higher metallicity.
We find equivalent results for $\amlt^{\odot}\left(\Delta s\right)$.
In addition, we find a tight correlation between the mixing-length
parameter and the inverse entropy jump. We derive an analytical expression
from the hydrodynamic mean-field equations that motivates the relation
to the mass mixing-length parameter, $\am$, and find that it qualitatively
shows a similar variation with stellar parameter (between $1.6$ and
$2.4$) with the solar value of $\am^{\odot}=1.83$. The vertical
correlation length scaled with the pressure scale height yields $1.71$
for the Sun, but only displays a small systematic variation with stellar
parameters, the correlation length slightly increases with $\teff$.}{We
derive mixing-length parameters for various stellar parameters that
can be used to replace a constant value. Within any convective envelope,
$\am$ and related quantities vary strongly. Our results will help
to replace a constant $\amlt$.}{}


\maketitle

\section{Introduction\label{sec:Introduction}}

In the past century, insights in various fields of physics led to
a substantially more accurate interpretation and understanding of
the processes taking place in the interior of celestial bodies. Astronomers
can parameterize the conditions on the surface of stars with theoretical
stellar atmosphere models, and with the theory of stellar structure
and evolution, they are additionally capable to predict the complex
development of stars.

The radiated energy of cool stars, originating from the deeper interior
because of nuclear burning in the center, is advected to the surface
by convective motions in the envelope that are driven by negative
buoyancy acceleration. At the thin photospheric transition region
the large mean free path of photons allows them to escape into space,
and the convective energy flux is abruptly released. To theoretically
model this superadiabatic boundary domain of stars is challenging
because of the nonlinear and nonlocal nature of turbulent subsurface
convection and radiative transfer, and an analytical solution is a
long-standing unresolved problem.

To account for the convective energy transport, \citet{BohmVitense:1958p4822}
formulated the mixing-length theory (MLT), which was initially proposed
by \citet{Prandtl:1925} in analogy to the concept of the mean free
path in the kinetic gas theory. In the framework of MLT, it is assumed
that the heat flux is carried by convective elements for a typical
distance before they dissolve instantaneously into the background.
This distance is the so-called mixing-length, $\lmlt$, usually expressed
in units of the pressure scale height, $\amlt=\lmlt/H_{P}$. The mixing-length
parameter $\amlt$ is a priori unknown, hence it has to be calibrated,
usually by matching the current radius and luminosity of the Sun by
a standard solar model with a single depth-independent $\amlt^{\odot}$.
This calibrated value for the Sun is then used for all stellar parameters.
We recall that $\amlt^{\odot}$, in fact, corrects for all other shortcomings
of the solar model, deficits in the equation-of-state (EOS), the opacities,
or the solar composition. It therefore is no wonder that its numerical
value \citep[typically around 1.7 to 1.9; e.g., see][]{Magic:2010p13816}
varies with progress in these aspects and from code to code. In addition,
MLT is a local and time-independent theory that effectively contains
three additional, free parameters, and assumes symmetry in the up-
and downflows, hence also in the vertical and horizontal direction.
The actual formulation of MLT can also vary slightly \citep[e.g., see][]{Henyey:1965p15592,Mihalas:1970p21310,Ludwig:1999p7606}. 

Many attempts have been made to improve MLT, a substantial one being
the derivation of a nonlocal mixing-length theory \citep{Gough:1977p8031,Unno:1985p22746,Deng:2006p22550,Grossman:1993p7953}.
The standard MLT is a local theory, meaning that the convective energy
flux is derived purely from local thermodynamical properties, ignoring
thus any nonlocal properties (e.g., overshooting) of the flow. Nonlocal
models are typically derived from the hydrodynamic equations, which
are a set of nonlinear moment equations including higher order moments.
To solve them, closure approximations are considered (e.g., diffusion
approximation, anelatistic approximations, or introducing a diffusion
length). Other aspects have also been studied: the asymmetry of the
flow by a two-stream MLT model \citep{Nordlund:1976p13639}, the anisotropy
of the eddies \citep{Canuto:1989p22737}, the time-dependence \citep{Xiong:1997p22562},
and the depth-dependence of $\amlt$ \citep{Schlattl:1997p1676}.
While standard MLT accounts for only a single eddy size (which is
$\lmlt$), \citet{Canuto:1991p6553} extended this to a larger spectrum
of eddy sizes by including the nonlocal second-order moment \citep{Canuto:1996p4779}.
The original Canuto-Mazzitelli theory -- also known as the\emph{ full
spectrum turbulence} model -- used the distance to the convective
region border as a proxy for the mixing-length; a later version \citep{Canuto:1992}
re-introduced a free parameter resembling $\amlt$.

These approaches are often complex, but so far, the standard MLT is
still widely in use, and a breakthrough has not been achieved, despite
all the attempts for improvements. In 1D atmosphere modeling, the
current procedure is to assume a universal value of $1.5$ for the
mixing-length parameter $\amlt$ \citep[see]{Gustafsson:2008p3814,Castelli:2004p4949}.
For full stellar evolution models, the solar ``calibration'' yields
values around $\sim1.7-1.9$ \citep[see, e.g.\ ]{Magic:2010p13816}.
Since the value of the mixing-length parameter sets the convective
efficiency and therefore changes the superadiabatic structure of stellar
models, an accurate knowledge of $\amlt$ for different stellar parameters
would be a first step in improving models in that respect. However,
apart from the Sun, other calibrating objects are rare and data are
much less accurate (see Sect.~\ref{sub:Comparison-with-observations}
for an example), such as binary stars with well-determined stellar
parameters. 

The mixing-length parameter can be deduced from multidimensional radiative
hydrodynamic (RHD) simulations, where convection emerges from first
principles \citep[e.g., see][]{Ludwig:1999p7606}. Over the past decades,
the computational power has increased and the steady development of
3D RHD simulations of stellar atmospheres has established their undoubted
reliability by manifold successful comparisons with observations \citep{Nordlund:1982p6697,Steffen:1989p18861,Ludwig:1994p18892,Freytag:1996p808,Stein:1998p3801,Nordlund:1990p6720,Nordlund:2009p4109}.
The 3D RHD models have demonstrated that the basic picture of MLT
is incorrect: there are no convective bubbles, but highly asymmetric
convective motions. Nonetheless, an equivalent mixing-length parameter
has been calibrated by \citet{Ludwig:1999p7606} based on 2D hydrodynamic
models by matching the resulting adiabats with 1D MLT models \citep[see ][for the metal-poor cases]{Freytag:1999p7637}.
The authors showed that $\amlt$ varies significantly with the stellar
parameters (from $1.3$ to $1.8$), and also studied the impact of
a variable $\amlt$ on a globular cluster \citep{Freytag:1999p7645}.
In addition, \citet{Trampedach:2007p5614} applied a grid of 3D atmosphere
models with solar metallicity to calibrate the mixing-length parameter
(from 1.6 to 2.0), and the so-called mass mixing-length \citep{Trampedach:2011p5920}.

In the present work we calibrate the mixing-length parameter with
a 1D atmosphere code that consistently employs the identical EOS and
opacity as used in the 3D RHD simulations (Sect.~\ref{sec:theoretical_models}).
We present the resulting mixing-length parameter in Sect.~\ref{sec:Mixing-length}.
We also determine the mass mixing-length -- the inverse of the logarithmic
derivative of the unidirectional mass flux -- in Sect. \ref{sec:Mass-mixing-length},
and the vertical correlation length of the vertical velocity (Sect.~\ref{sec:velocity_correlation_length})
directly from the 3D atmosphere models. For the former quantity, we
derive a relation from the hydrodynamic mean-field equations that
demonstrates the relation to $\amlt$, which is further substantiated
by our numerical results. Finally, we conclude in Sect.~\ref{sec:Conclusions}.

\section{Theoretical models\label{sec:theoretical_models}}

\begin{figure*}
\includegraphics[width=88mm]{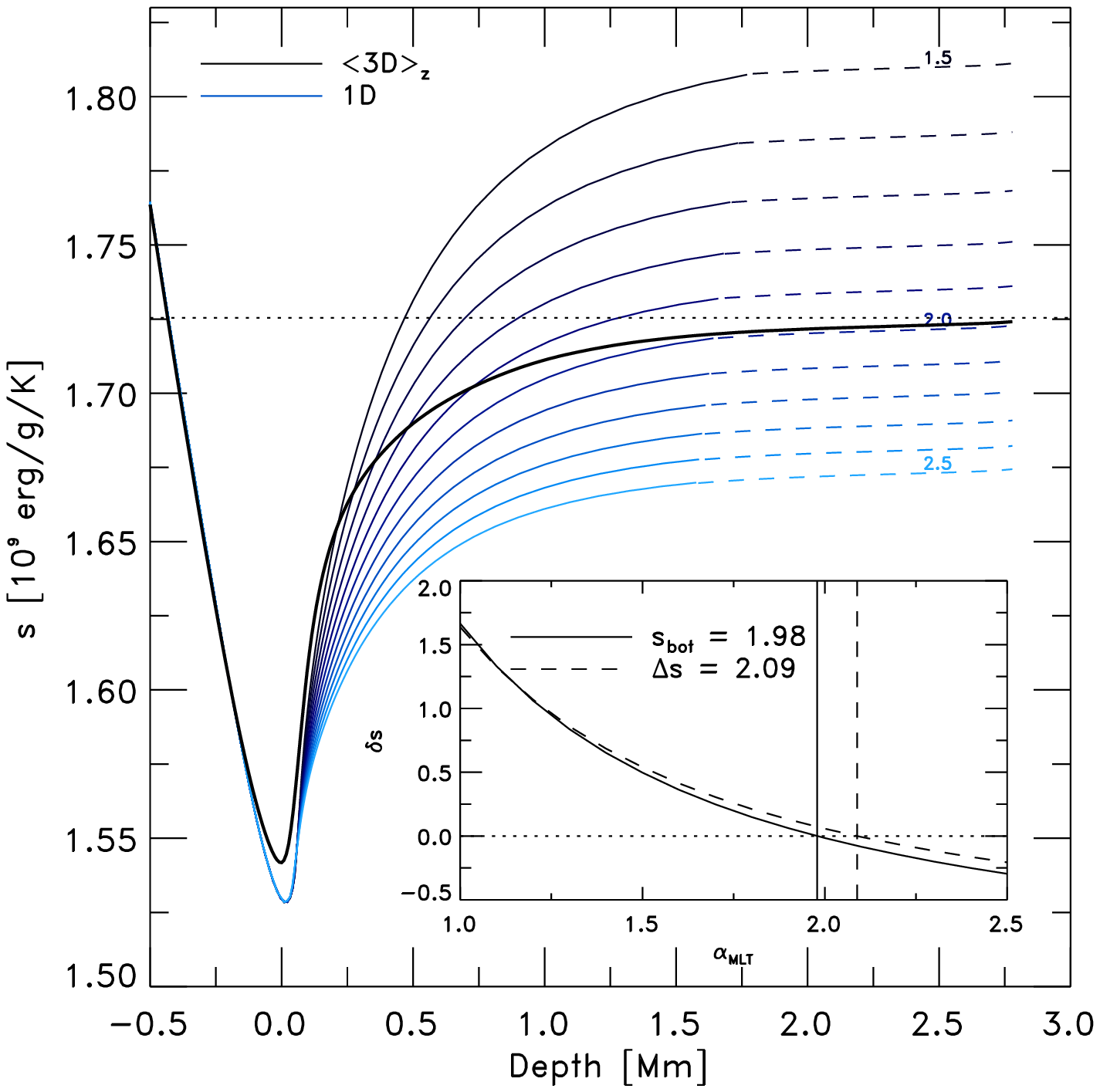}\includegraphics[width=88mm]{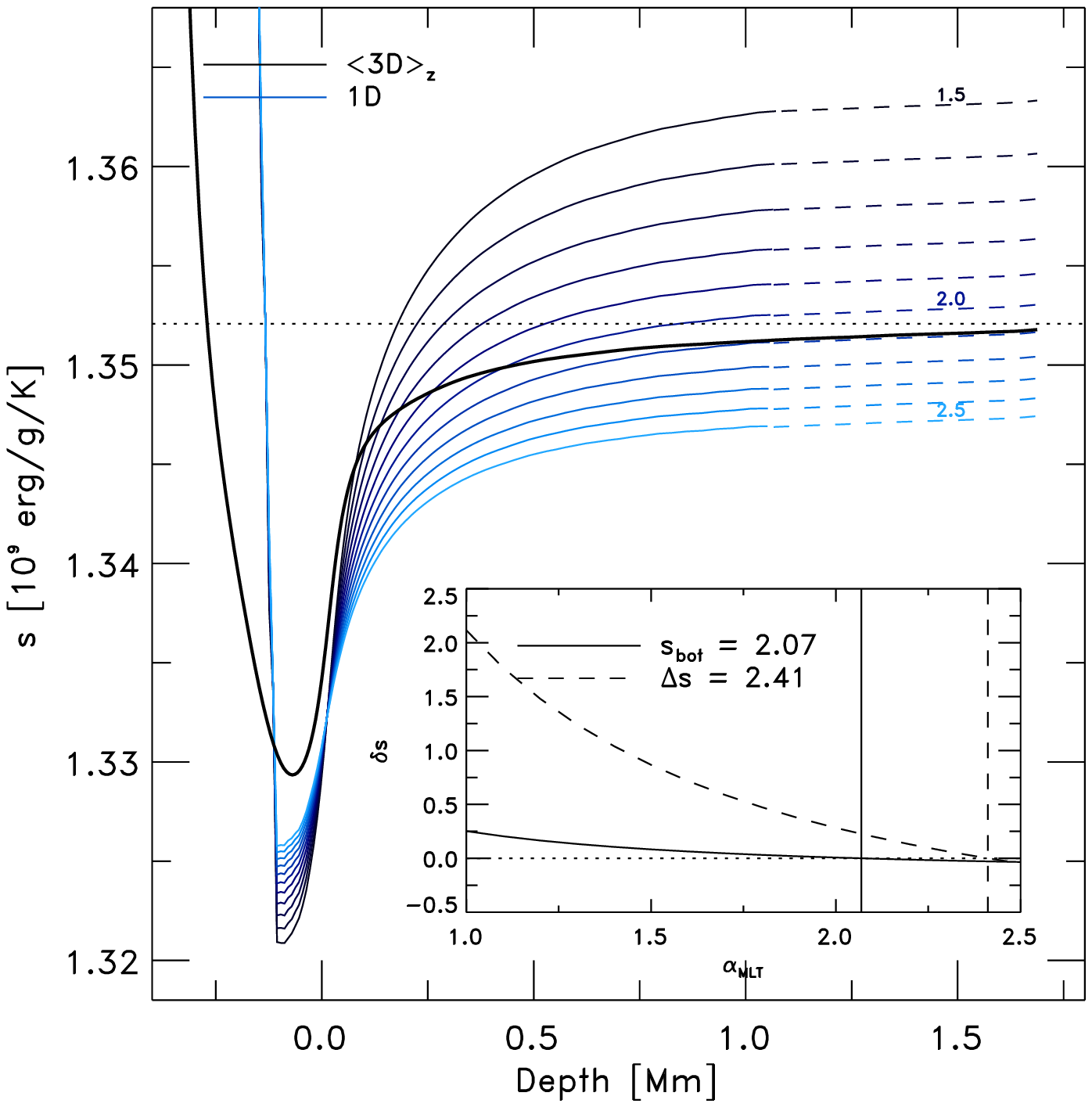}

\caption{\label{fig:match_alpha_mlt_atmo}Mean $\havz$ entropy (\emph{black
solid line}) vs.\ depth, and 1D models for different mixing-length
parameters, $\amlt=1.5-2.5$ (\emph{blue lines}), for the solar model
(\emph{left panel}) and a metal-poor dwarf with $\feh-2.0,\,\teff=4500\,\mathrm{K}$,
and $\logg=4.5$ (\emph{right panel}). We indicate the constant entropy
value of the deep adiabatic convection zone, $\ssbot$, in both figures
by the horizontal dotted line. In the deeper layers, we extended the
1D models (\emph{dashed lines}) with the aid of the entropy gradient
from the $\hav$ models. The calibration of the mixing-length parameter
$\amlt$ for the solar model is illustrated by the smaller insets,
which depict the relative differences between the 1D and 3D models
($\delta s=s_{1\mathrm{D}}/s_{3\mathrm{D}}-1$) for $\ssbot$ (\emph{solid})
and the entropy jump $\Delta s$ (\emph{dashed}). For the solar model
the two approaches result in $\amlt=1.98$ and $2.09$.}
\end{figure*}

\subsection{3D atmosphere models\label{sub:3D-atmosphere-models}}

We computed the \textsc{Stagger}-grid, a large grid of 3D RHD atmosphere
models that covers a wide range in stellar parameter space \citep[see][hereafter Paper I]{Magic:2013}.
The 3D atmosphere models were computed with the \textsc{Stagger}-code,
which solves the 3D hydrodynamic equations for conservation of mass,
momentum and energy, coupled with a realistic treatment of the radiative
transfer. We employed the EOS by \citet{Mihalas:1988p20892}, and
up-to-date continuum and line opacities \citep{Gustafsson:2008p3814}.
For the solar chemical abundances, we used the values by \citet[hereafter AGS09]{Asplund:2009p3308}.
Our simulations are of the so-called box-in-a-star\emph{ }type, that
is we compute only a small, statistically representative volume that
includes typically ten granules. Our (shallow) simulations only cover
a small fraction of the total depth of the convective envelope. Because
of the adiabaticity of the gas in the lower parts of the simulation
box, the asymptotic entropy value of the convective zone, $s_{\mathrm{ad}}$,
is matched by the fixed entropy at the bottom of the simulation domain,
$\ssbot$, which is one of the simulation parameters. The effective
temperature is therefore a result in our 3D simulations, and is actually
a temporally averaged quantity. In 1D models $\teff$ is an actual
fixed input value in addition without fluctuations. 

We determine the entropy jump, $\Delta s$, as the difference between
the entropy minimum and the constant entropy value of the adiabatic
convection zone with $\Delta s=s_{\mathrm{min}}-\ssbot$. In \citet[hereafter Paper II]{magic:2013arXiv1307.3273M},
we studied in detail the differences between mean $\hav$ models resulting
from different reference depth scales. In the present work, we show
and discuss only averages on constant geometrical height $\havz$,
since these fulfill the hydrodynamic equilibrium and extend over the
entire vertical depth of the simulations. The \textsc{Stagger}-grid
encompasses $\sim220$ models ranging in effective temperature, $T_{\mathrm{eff}}$,
from $4000$ to $7000\,\mathrm{K}$ in steps of approximately $500\,\mathrm{K}$
(we recall that $T_{\mathrm{eff}}$ is the result of the input quantity
$s_{\mathrm{ad}}$, and the intended $T_{\mathrm{eff}}$ grid point
values are adjusted within a margin below 100~K). The surface gravity,
$\log g$, ranges from $1.5$ to $5.0$ in steps of 0.5 dex, and metallicity,
$\left[\mathrm{Fe}/\mathrm{H}\right]$, from $-4.0$ to $+0.5$ in
steps of $0.5$ and $1.0\,\mathrm{dex}$. We refer to Paper~I for
detailed information on the actual methods for computing the grid
models, their global properties, and mean stratifications.

\subsection{1D atmosphere models\label{sub:1D-atmosphere-code}}

For the \textsc{Stagger}-grid, a 1D MLT atmosphere was developed that
uses exactly the same opacities and EOS as the 3D models (Paper I).
Therefore, the chemical compositions are identical. The code uses
the MLT formulation by \citet{Henyey:1965p15592} (see App. \ref{app:Mixing-length-formulation}
for details), similar to the MARCS code \citep{Gustafsson:2008p3814}.
Furthermore, for consistency, the 1D models were computed with exactly
the same $\teff$ as the 3D models.

The actual implementation of MLT differs slightly depending on the
considered code \citep[e.g.,][]{Ludwig:1999p7606}. In the standard
MLT formulation there are four parameters in total. The mixing-length
parameter, $\amlt=\lmlt/H_{p}$, sets the convective efficiency, while
$y=3/(4\pi^{2})\simeq0.076$ is assumed for the temperature distribution,
and $\nu=8$ for the turbulent viscosity (see App. \ref{app:Additional-MLT-parameters}
for a discussion). We only considered the mixing-length parameter
$\amlt$ for the calibration, while the additional parameters were
kept fixed to their default values, and the turbulent pressure was
entirely neglected.

\section{mixing-length parameter\label{sec:Mixing-length}}

\begin{figure*}
\includegraphics[width=88mm]{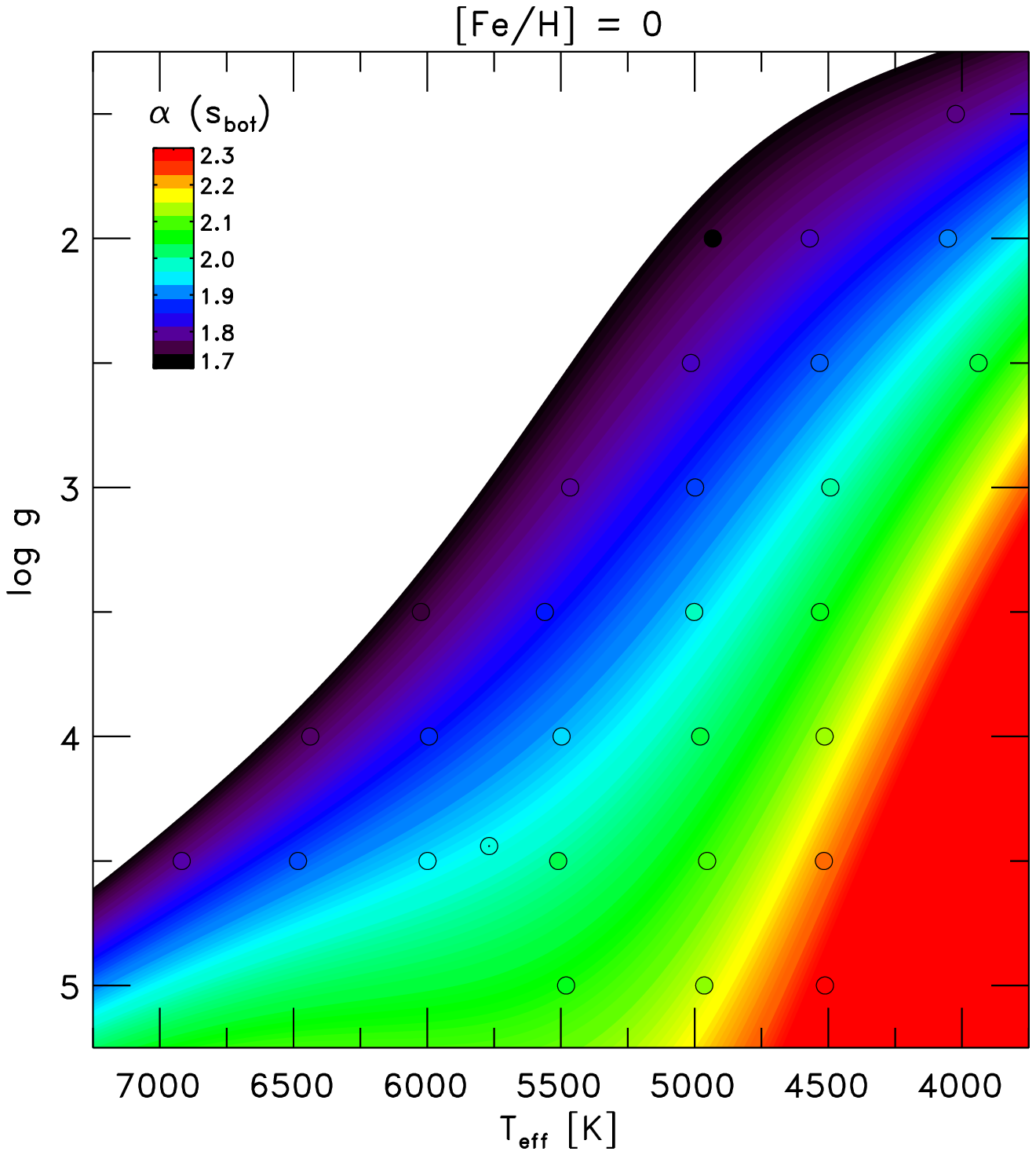}\includegraphics[width=88mm]{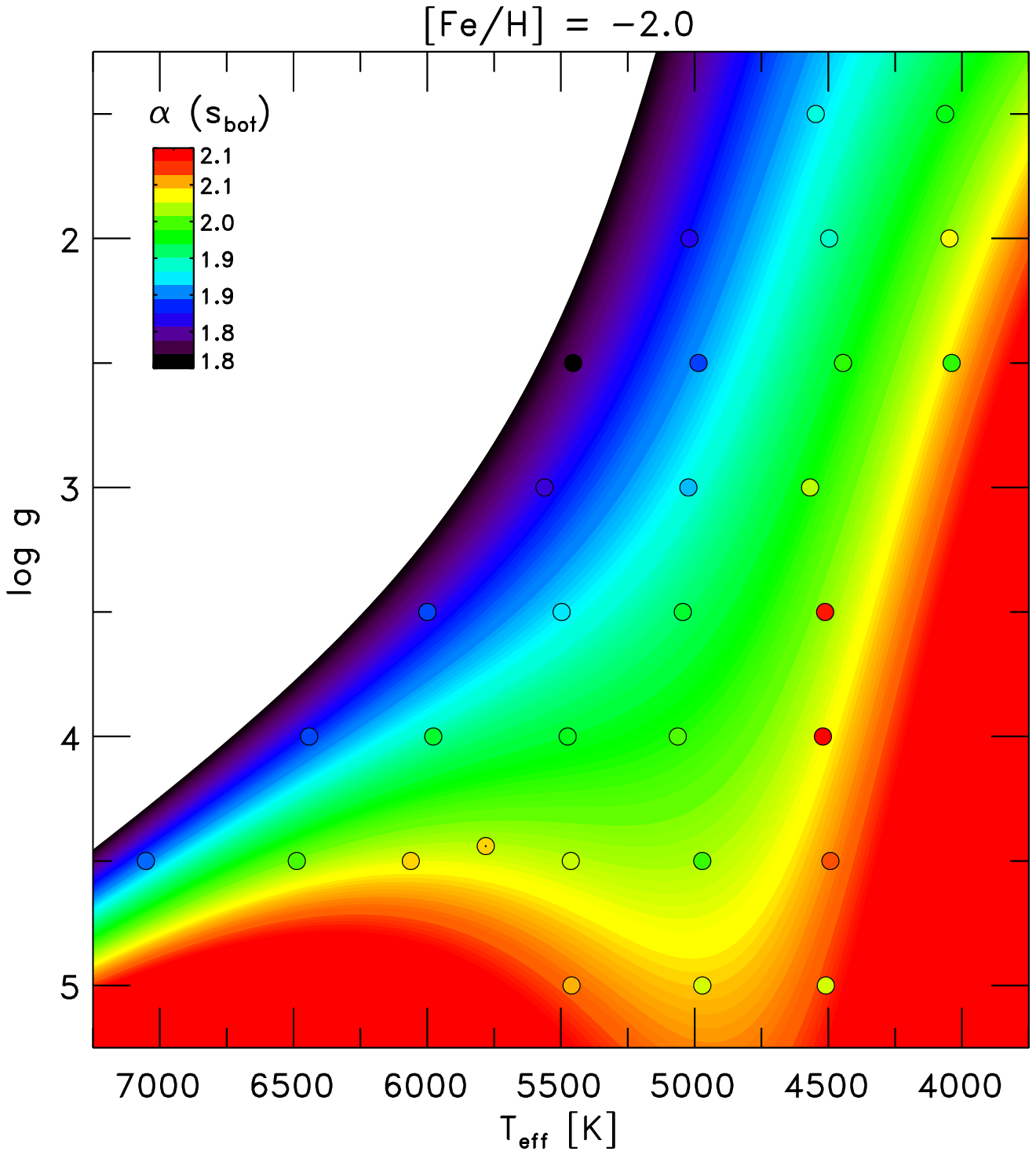}

\caption{\label{fig:alpha_mlt_atmo}Kiel-diagram ($\teff-\logg$ diagram) with
the mixing-length parameter calibrated with the constant entropy value
of the adiabatic convection zone, $\amlt\left(\ssbot\right)$, for
solar and subsolar metallicity (\emph{left} and \emph{right panels},
respectively). The mixing-length is color-coded as indicated and shown
with contours derived from functional fits (see App. \ref{app:Functional-fits}),
while the circles represent the \textsc{Stagger}-grid models. Note
the difference in the color scales.}
\end{figure*}
\begin{figure*}
\includegraphics[width=88mm]{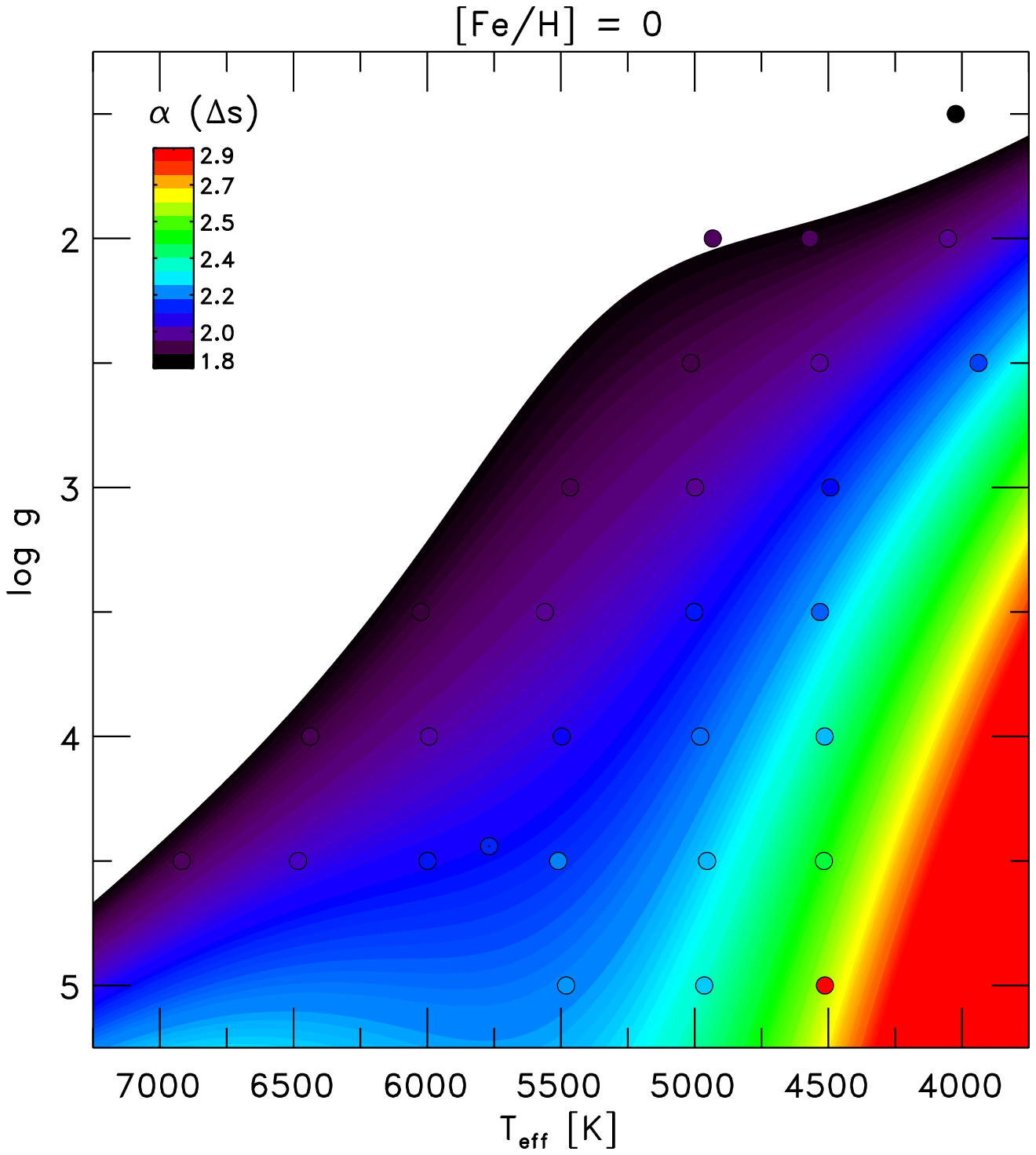}\includegraphics[width=88mm]{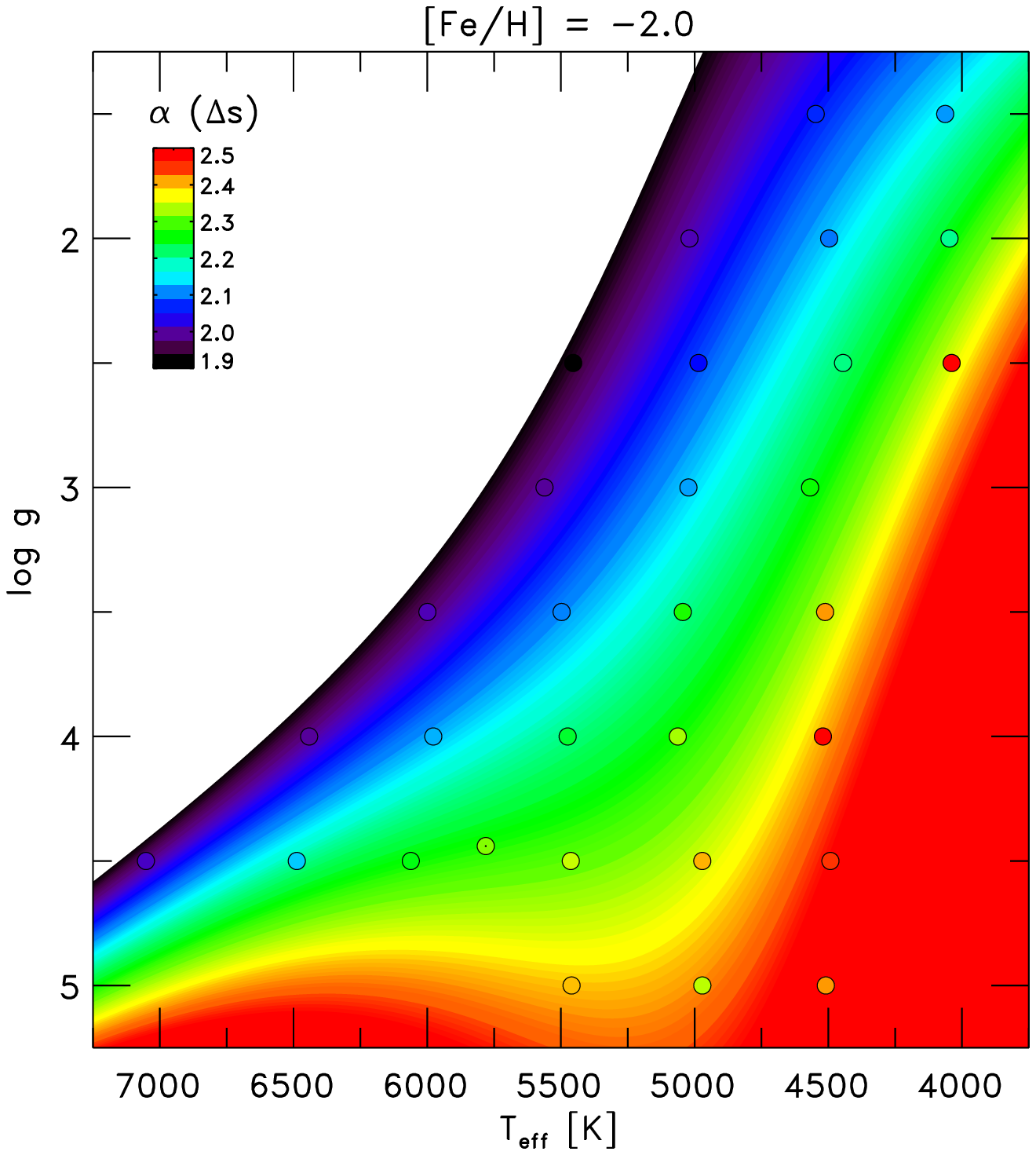}

\caption{\label{fig:alpha_mlt_atmo_jump}As Fig. \ref{fig:alpha_mlt_atmo},
but here we show the mixing-length parameter calibrated with the entropy
jump $\amlt\left(\Delta s\right)$.}
\end{figure*}
\begin{figure*}
\includegraphics[width=176mm]{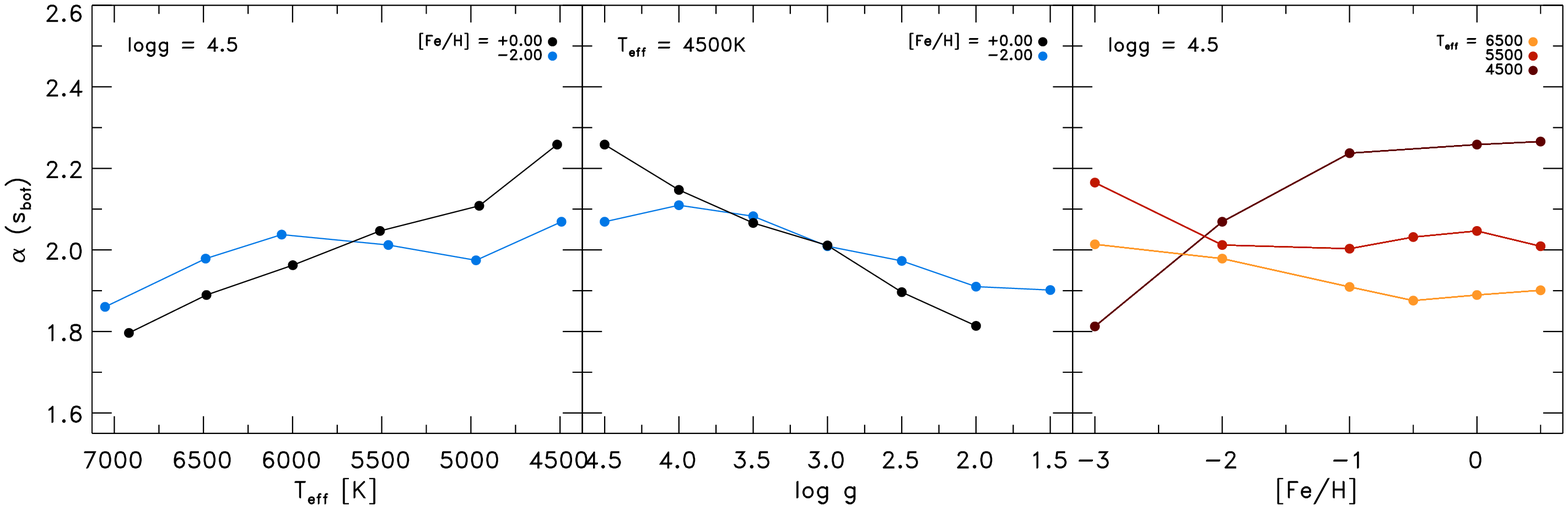}

\includegraphics[width=176mm]{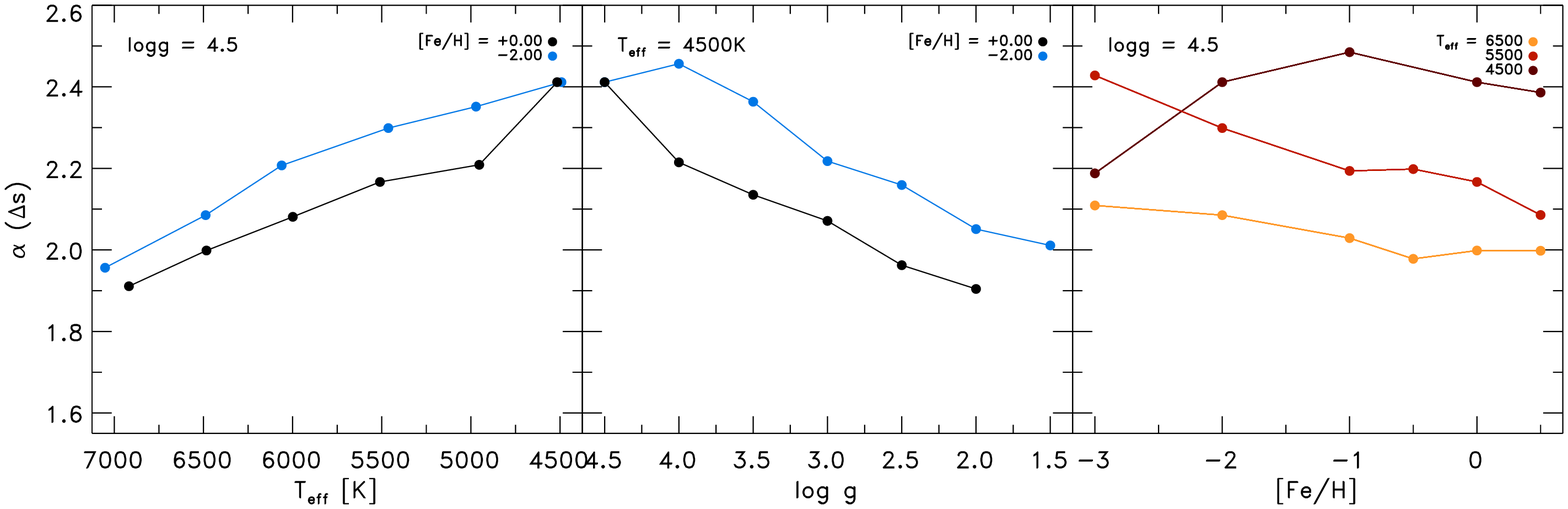}

\caption{\label{fig:alpha_mlt_stellar_parameter}Dependence of the mixing-length
parameters on the different stellar parameters ($\amlt\left(\ssbot\right)$
and $\amlt\left(\Delta s\right)$ in the \emph{top} and \emph{bottom}
panel, respectively). We varied one stellar parameter at a time, while
the other two were kept fixed (\emph{left}: effective temperature;
\emph{middle}: surface gravity; \emph{right}: metallicity). The fixed
stellar parameters are indicated and color-coded.}
\end{figure*}

\subsection{Matching the mixing-length parameter\label{sub:Matching-the-mixing-length}}

We calibrated $\amlt$ by matching either the asymptotic entropy value
of the deep convection zone, $\ssbot$, or the entropy jump, $\Delta s$,
from the 1D and 3D models. We refer to these throughout as $\amlt\left(\ssbot\right)$
and $\amlt\left(\Delta s\right)$. The value of $\ssbot$ is an input
parameter in our 3D simulations and represents the adiabatic entropy
of the incoming upflows at the bottom of the box that are replenishing
the outflows. The horizontally and temporally averaged entropy at
the bottom, $\left\langle s\right\rangle _{\mathrm{bot}}$, in contrast,
considers both the up- and downflow, and is thus slightly lower than
$\ssbot$ because of the entropy-deficient downflows. However, in
our simulations the deeper layers have almost adiabatic conditions.
The contrast of the thermodynamic variables at the bottom is extremely
low with $\left\langle X\right\rangle _{\mathrm{bot}}-X_{\mathrm{bot}}\ll1\,\%$.

For the calibration, we computed 1D models with $\amlt$ from $1.0$
to $2.5$ in steps of $0.1$ and determined $\amlt$ by minimizing
the difference $\delta s=\ssbot^{\mathrm{1D}}-\ssbot^{\mathrm{3D}}$
or the difference in the entropy jumps $\delta s=\Delta s^{1\mathrm{D}}-\Delta s^{3\mathrm{D}}$.
We remark that some 1D atmosphere models had convergence problems,
when they were extended to the same depth as the 3D models. Therefore,
we had to calculate slightly shallower 1D models. However, we extended
the 1D entropy stratifications with the entropy gradients of the $\hav$
model (see Fig. \ref{fig:match_alpha_mlt_atmo}). We performed tests
by truncating $\hav$ models and extending them with our method, which
led to the same stratification. Therefore, we assume that the missing
depth in the 1D entropy run leads to only minor uncertainties in the
resulting $\amlt$. We fitted the differences, $\delta s$, with a
second-order polynom to derive the value of $\amlt$. We emphasize
that the calibration of $\amlt$ is more meaningful for identical
EOS, and the entropy is consistently computed. For the calibration,
we neglected the turbulent pressure in the 1D models entirely (i.e.,
$\beta=0$). Including turbulent pressure would clearly influence
the calibration of $\amlt$, but, to account properly for $p_{\mathrm{turb}}$,
one would need to employ an improved description of convection that
accounts for nonlocal effects \citep[private communication with D. Gough, and see also][]{Ludwig2008IAUS..252...75L}.
Because of the local nature of the standard MLT, the impact of the
turbulent pressure is confined to the convective region and the turbulent
leviation is rendered poorly. We note that the influence of the turbulent
pressure is included in the calibrated $\amlt$ values.

In Fig.~\ref{fig:match_alpha_mlt_atmo}, we illustrate the calibration
for the solar model and for a cool metal-poor dwarf with the mean
entropy, $s$, in the convection zone. For the solar simulation, we
determined a mixing-length parameter of $\amlt=1.98$ and $2.09$
from matching either the adiabatic entropy value (left panel) or the
entropy jump (right panel). Note how $s$ converges asymptotically
against $\ssbot$. Furthermore, it is also evident from Fig.~\ref{fig:match_alpha_mlt_atmo}
that for a higher $\amlt$ the adiabat ($\ssbot$) of the 1D models
is decreasing in the convection zone. The entropy minimum of the $\havz$
on geometrical height is slightly mismatched by the 1D models, which
is reflected by slightly different calibrated $\amlt\left(\Delta s\right)$
values. In the 1D models $s_{\mathrm{min}}$ varies only little for
different $\amlt$, and the differences, $\Delta\amlt$, are between
$\sim10^{-4}$ and $10^{-3}$ (cf.\ also the right panel). Since
the entropy jumps are in general much larger than the variation of
$s_{\mathrm{min}}$, their influence is very weak, and only for very
cool metal-poor models with very small entropy jumps, differences
in $s_{\mathrm{min}}$ influence the calibration (see right panel
in Fig. \ref{fig:match_alpha_mlt_atmo}).

We find in general very similar results for $\amlt$ by employing
a 1D \emph{envelope} code, which solves the stellar structure equations
down to the radiative interior by including the same EOS and opacities
\citep{Christensen-Dalsgaard:2008Ap&SS.316...13C}. This is in particular
true for solar metallicity. The 1D envelope code relies on an assumed
$\ttau$ relation in the (Eddington gray) atmosphere, which obviously
influences the thermal stratification at the outer boundary of the
convective envelope. In particular, metal-poor 1D convective interior
models with a fixed $\ttau$ relation are affected by this, and will
return different mixing-length parameters. The 1D atmosphere code
works without the need for any $\ttau$ relation, since it solves
the radiative transfer by itself. We therefore present and discuss
only the mixing-length parameters matched by the 1D atmosphere code.

Furthermore, we have performed functional fits for the calibrated
mixing-length parameters, that is $\amlt=f\left(\teff,\logg,\feh\right)$,
with the same functional basis as used by \citet{Ludwig:1999p7606}.
For more details see App. \ref{app:Functional-fits}, and the resulting
coefficients are provided in Table \ref{tab:hgl_fit}. In Table \ref{tab:hgl_fit}
we also listed the uncertainties of the fits estimated with the root-mean-square
and highest deviation, which are increasing for lower metallicities.

\subsection{Calibrations with the adiabatic entropy value\label{sub:matching_ssbot_atmo}}

In Fig.~\ref{fig:alpha_mlt_atmo}, we show an overview of the variation
of the $\amlt$ values calibrated with $\ssbot$ for different stellar
parameters in the Kiel-diagram, in particular, for two illustrating
metallicities ($\mathrm{[Fe/H]}=0$ and $-2$). The mixing-length
parameter varies rather systematically in the range between $\sim1.7$
and $\sim2.3$: $\amlt$ increases for lower $\teff$ and $\feh$
and higher $\logg$ (see also Fig. \ref{fig:alpha_mlt_stellar_parameter}).
Towards lower metallicity, models with cooler $\teff$ deviate from
a linear run, which can be attributed to the differences in the outer
boundary condition of the 1D models. A larger $\amlt$ relates to
a higher convective efficiency, which implies that a smaller entropy
jump is necessary to carry the same convective energy flux. Indeed,
we find the entropy jump to increase for higher $\teff$, lower $\logg$,
and higher $\feh$ (see Paper I); we find that $\amlt$ varies qualitatively
inversely to the entropy jump. The mixing-length parameter is inversely
proportional to the variation of the logarithmic values of the entropy
jump, the peak in the entropy contrast, and vertical rms-velocity
(see Sect. \ref{sub:Comparison-with-global}). This agrees with the
fact that both the entropy jump and the mixing-length parameter are
related to the convective efficiency (see Sect. \ref{sub:Comparison-with-global}).

\subsection{Calibrations with the entropy jump\label{sub:matching_ssj_atmo}}

\begin{figure}
\includegraphics[width=88mm]{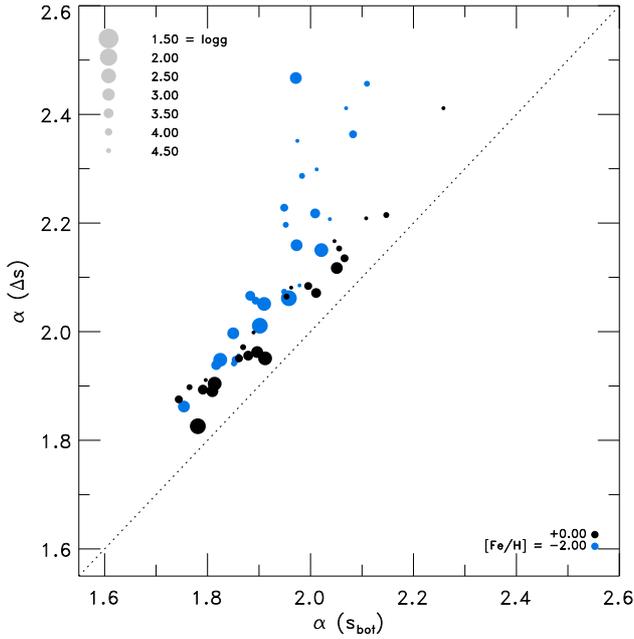}\caption{\label{fig:alpha_ssbot_ssj}Comparison of the mixing-length parameter
calibrated with the entropy jump, $\amlt\left(\Delta s\right)$, and
the constant entropy value of the adiabatic convection zone, $\amlt\left(\ssbot\right)$,
for different stellar parameters.}
\end{figure}
We also calibrated the mixing-length parameter with the 1D MLT atmosphere
code by matching the entropy jump $\Delta s$. The resulting values
are summarized for two metallicities in Fig.~\ref{fig:alpha_mlt_atmo_jump},
showing a similar behavior as the results of the previous section
(see also Fig. \ref{fig:alpha_mlt_stellar_parameter}). We find that
the $\amlt$ values based on $\Delta s$ are systematically higher
by $\sim0.1$ (between $\sim1.8$ and $\sim2.4$) than the values
based on $\ssbot$ (Fig.~\ref{fig:alpha_ssbot_ssj}), but the range
in $\amlt\left(\Delta s\right)$ is with $\Delta\amlt\approx0.6$
very similar to that for $\amlt\left(\ssbot\right)$. The differences
arise from the minimum of the entropy $s_{\mathrm{min}}$ around the
optical surface, which is lower for the 1D models than for the $\hav$
model (see Fig.~\ref{fig:match_alpha_mlt_atmo}), and therefore leads
to larger mixing-length parameters. The metal-poor simulations deviate
more strongly between $\amlt\left(\Delta s\right)$ and $\amlt\left(\ssbot\right)$,
since the boundary effect, induced by the differences in $\Delta s$,
increases for lower $\feh$. We note that the entropy jump is a relative
value, and consequently, the matching is less sensitive to outer boundary
effects.

\subsection{Comparison with global properties\label{sub:Comparison-with-global}}

\begin{figure}
\includegraphics[width=88mm]{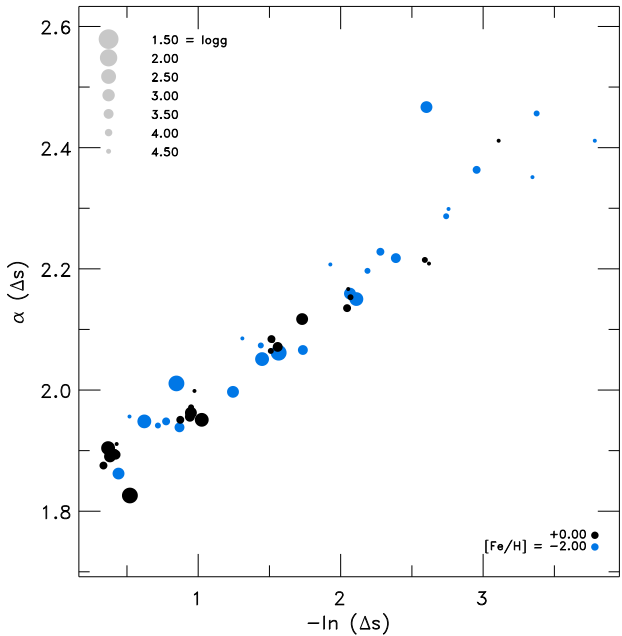}\caption{\label{fig:alpha_mltj_ssj}Comparison between the mixing-length parameter
calibrated with the entropy jump, $\amlt\left(\Delta s\right)$, and
the logarithm of the inverse of the entropy jump, $-\ln\left(\Delta s\right)$,
for different stellar parameters.}
\end{figure}
\begin{figure}
\includegraphics[width=88mm]{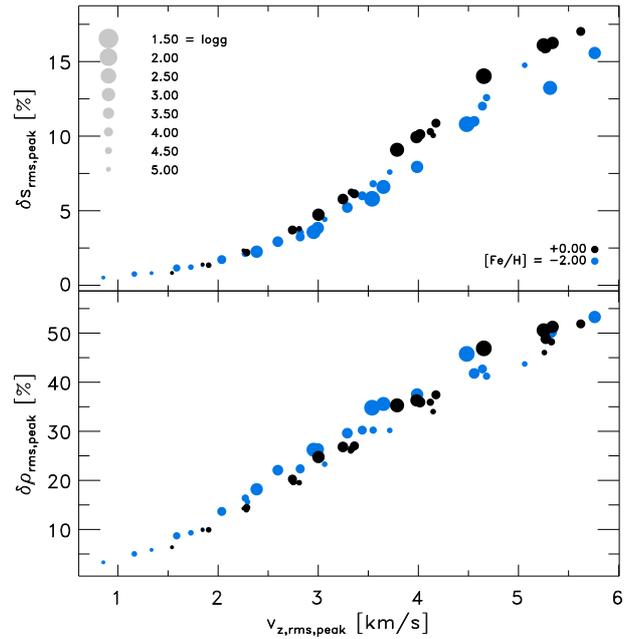}

\caption{\label{fig:psg_ssc_rhoc_uyrms}Highest contrast of the entropy and
density compared with the highest vertical rms-velocity (top and bottom
panel, respectively) for different stellar parameters.}
\end{figure}
We searched for systematic correlations between the mixing-length
parameter and mean thermodynamic properties. The inverse of the entropy
jump correlates well with $\amlt$. In Fig.~\ref{fig:alpha_mltj_ssj}
we demonstrate this by comparing the mixing-length parameter $\amlt\left(\Delta s\right)$
with the logarithm of the inverse of the entropy jump. Convection
is driven by radiative cooling in the surface layers. The entropy
jump results from the radiative losses at the optical surface, therefore,
the correlation of $\amlt$ originates in the interplay of the opacity,
$\kappa_{\lambda}$, radiative cooling rates, $q_{\mathrm{rad}}$,
and vertical velocity, $\vzrms$. The vertical velocity results from
buoyancy forces, $f_{b}=g\Delta\rho$, that act on the overturning,
overdense flows at the optical surface.\textcolor{blue}{{} }Hence, a
larger entropy jump will entail higher contrast in the entropy and
density ($\delta s_{\mathrm{rms}}$ and $\delta\rho_{\mathrm{rms}}$),
which will induce a stronger downward acceleration. We illustrate
this in Fig.~\ref{fig:psg_ssc_rhoc_uyrms}, where the peak values
for $\delta s_{\mathrm{rms}}$ and $\delta\rho_{\mathrm{rms}}$ in
the superadiabatic region are plotted against the peak vertical rms-velocity.
Evidently, the entropy and density contrast correlate well with the
vertical velocity, and this is the underlying reason for the tight
(inverse) correlation between mixing-length parameter and entropy
jump. In Paper I we have already discussed the correlation of the
entropy jump with the peak vertical velocity and the density at the
same location, and we deduced the reason for this in the convective
energy flux, which essentially contains these quantities.

\subsection{Comparison with 2D calibrations\label{sub:Comparison-with-2d}}

\begin{figure}
\includegraphics[width=88mm]{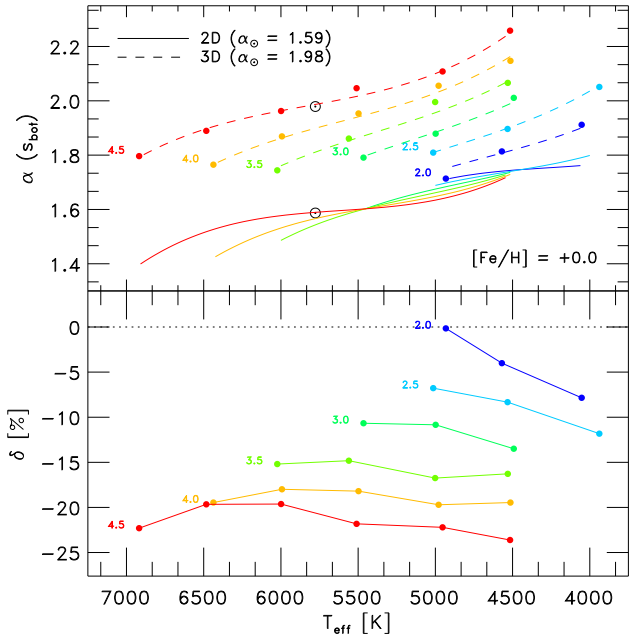}\caption{\textbf{\label{fig:comp_alpha_2d}}\emph{Top panel}: calibration of
$\amlt$ with 2D \citep{Ludwig:1999p7606} and 3D simulations in comparison
(\emph{solid} and \emph{dashed lines}, respectively). The surface
gravity is indicated and color-coded. The solar values are indicated.
\emph{Bottom panel}: relative differences ($\delta=2\mathrm{D}/3\mathrm{D}-1$).}
\end{figure}
We compared the differences between our inferred mixing-length parameters
with those of \citet{Ludwig:1999p7606} based on similar, but 2D hydrodynamical
surface convection simulations. These authors also matched the resulting
2D-based $\ssbot$ by varying $\amlt$ of a 1D envelope code that
uses the same EOS and opacity. However, the EOS and opacity are not
identical to those used by us, and there are other differences in
the models, such as, most importantly, the solar composition. This
needs to be kept in mind when interpreting the comparison.

We also remark that \citet{Ludwig:1999p7606} derived $\ttau$-relations
from the 2D models, and used them for the 1D models as boundary conditions
to render the entropy minimum of the 2D simulations more correctly.
In Paper~I we noticed that $\ssbot$ resulting from the \textsc{Stagger}-grid
is very similar to values from the 2D grid, while the entropy jump
$\dss$ differs slightly.

In Fig.~\ref{fig:comp_alpha_2d} we compare the calibrated mixing-length
parameter from both studies. The results of \citet{Ludwig:1999p7606}
also show a clear $\teff$-dependence, while surface gravity has only
very little influence on $\amlt$. While the 3D-calibrated mixing-length
parameter decreases with lower surface gravity, its 2D equivalent
increases moderately. Their solar mixing-length parameter is $\amlt=1.59$,
which is lower by 0.39 ($\sim20\%$) than our mixing-length parameter,
but similar to the solar model value of that time, as is ours for
the present generation of solar models. The $\amlt$ values for dwarf
models ($\logg=4.5$) are in general around $20\,\%$ lower than in
our case. Towards giants the difference decreases, since the 3D values
decrease with $\log g$. For 3D convection simulations it is known
that convection is more efficient than for the 2D case. Therefore,
the mixing-length parameters derived from the 3D models are in general
systematically larger. Taking into account the model generation effect,
the comparison is quite satisfactory with the exception of the discrepant
$\log g$-dependence.

\subsection{Impact on stellar evolutionary tracks\label{sub:Impact-on-tracks}}

\begin{figure}
\includegraphics[width=88mm]{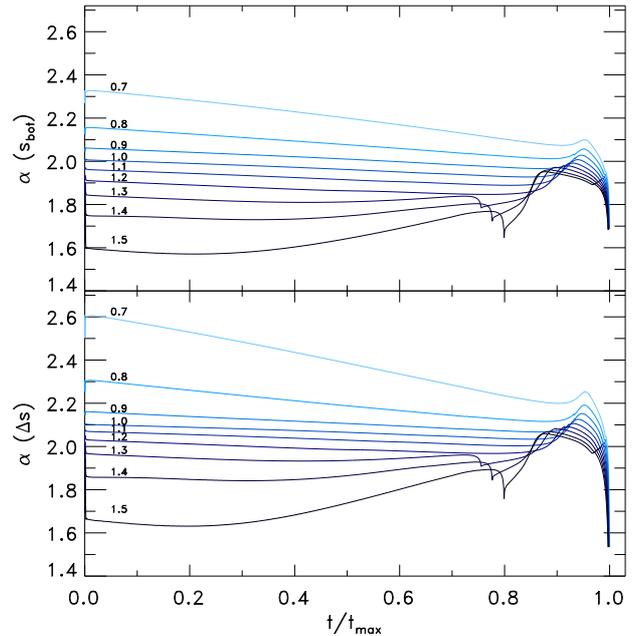}

\caption{\label{fig:tracks_alpha}Mixing-length parameter along stellar evolutionary
tracks with solar metallicity against the normalized age for the masses
from 0.7 to 1.5 $M_{\odot}$ (indicated). The tracks are derived from
the functional fits $f\left(\teff,\logg\right)$ of $\amlt$-calibrations
with $\ssbot$ and $\dss$ (\emph{top} and \emph{bottom panel}, respectively)
and all tracks end on the RGB when $\logg=1$.}
\end{figure}
The variation of $\amlt$ along typical stellar evolutionary tracks
ranges from $1.6$ to $2.4$ from higher to lower mass (see Fig.~\ref{fig:tracks_alpha}),
and deviates by up to $\pm20\,\%$ from the solar value ($\amlt^{\odot}$).
Note that in Fig. \ref{fig:tracks_alpha} we show $\amlt$ along tracks
calculated with a constant value of the mixing-length parameter ($1.78$)
obtained from the usual solar model calibration \citep[see][]{Magic:2010p13816}.
The figure therefore does not show the actual, self-consistent changes
in $\amlt$ along the evolution, but, significant differences are
hardly to be expected. During the main-sequence evolution $\amlt$
varies only little and is almost constant, in particular for the lower
masses without a convective core. The variable mixing-length parameter
has a stronger influence during later evolutionary stages, the TO
and the RGB ascent; $\amlt$ increases first towards values around
$\sim1.9-2.1$, and then drops sharply to values of $\sim1.7$ for
all masses, which is the consequence of the narrow range in red giant
temperature and surface gravity.

The mixing-length parameter not only determines $T_{\mathrm{eff}}$
of the stellar models, but also influences the adiabatic stratification
of the 1D models in the deeper convection zone. In particular, for
a larger $\amlt$ the lower boundary of the convection zone is located
deeper in the interior. Therefore, for stars with lower (higher) masses,
a variable mixing-length parameter with stellar parameter will increase
(decrease) the depth of the convection zone. As a consequence, one
can expect that the convective mixing will be enhanced (reduced) for
less (more) massive stars in stellar evolutionary calculations. This
may influence, for example, the depletion and burning of Li in low-mass
stars.

\subsection{Comparison with observations\label{sub:Comparison-with-observations}}

\begin{figure}
\includegraphics[width=88mm]{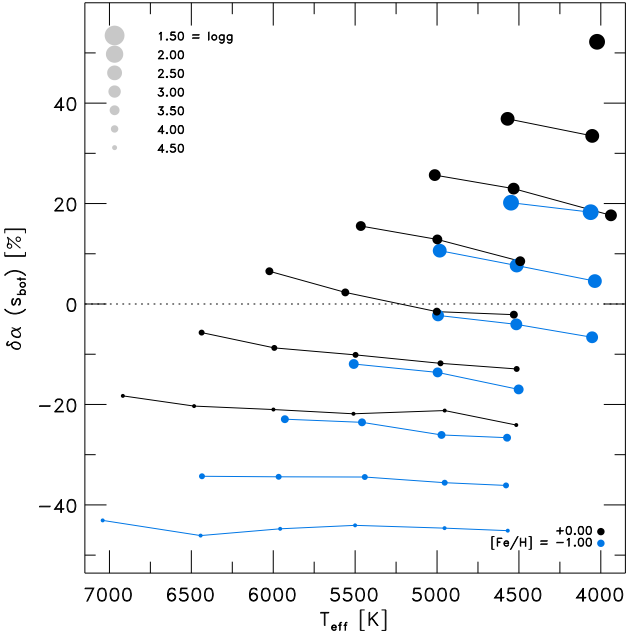}\caption{\label{fig:comp_alpha_observation}Relative differences between the
the mixing-length parameter derived from observations and our 3D RHD
models for different stellar parameters.}
\end{figure}
Observations provide an opportunity to constrain free parameters in
theoretical models. \citet{Bonaca:2012p19742} attempted to calibrate
the mixing-length parameter from Kepler-observations of dwarfs and
subgiants (90 stars). Employing the usual scaling relations for the
frequency of the maximal oscillation mode power, $\nu_{\mathrm{max}}$,
and the large frequency separation, $\Delta\nu$ \citep[see, for example,][]{huber:2011},
in connection with $\teff$ and $\feh$ from spectroscopic observations,
they estimated mass and radius of the observed objects. From a grid
of stellar evolutionary tracks computed with different $\amlt$ values,
they then selected the value that matched the inferred stellar parameters.
The stellar evolutionary tracks were computed with the Yale stellar
evolution code by employing the EOS and opacities from OPAL \citep[see][]{Demarque:2008p17031}.
For the outer boundary conditions they used the Eddington $\ttau$
relation and the standard MLT formulation by \citet{BohmVitense:1958p4822}.
These differences need be considered in the comparison of the resulting
$\amlt$ values.

\citet{Bonaca:2012p19742} derived an average mixing-length parameter
of $1.60$ from the observations, which is in general lower than their
solar-calibrated value with 1.69, which resulted from the 1D models
without the comparison with observations. We compare the (linear)
functional fit of $\amlt$ derived in \citet{Bonaca:2012p19742},
with stellar parameters to our own results in Fig.~\ref{fig:comp_alpha_observation}.
We compare the calibration resulting from their complete data set.
They also derived a fit for a subset of dwarfs, which is quite restricted
in the range of stellar parameters and quite different from the fit
for the full sample, however, they determined the solar mixing-length
parameter with $\amlt^{\odot}=1.59$, which is $20\,\%$ lower than
our result of $1.98$. However, we remark that because of differences
in the input physics and methods, the comparison between absolute
values of $\amlt$ is limited. Interestingly, the variation with $\teff$
for a given $\logg$ and $\feh$ is rather similar apart from an almost
constant offset. For different $\logg$ and $\feh$, we find significant
systematical differences (see Fig. \ref{fig:comp_alpha_observation}).
The values for dwarfs are in general lower by up to $\sim20-40\,\%$
depending on gravity and metallicity, while the values for giants
are greater by the similar amount. The comparison is made more difficult
because even the full sample of \citet{Bonaca:2012p19742} is rather
limited in $\logg$, and biased towards dwarfs. Additionally, the
input physics (EOS and opacity) of their models deviates from ours.
The authors themselves mention the absence of strong correlations
with $\log g$, their restricted range in $\mathrm{[Fe/H]}$, the
discrepancies to the results by \citet{Ludwig:1999p7606} and \citet{Trampedach:2007p5614},
and the fact that $\amlt$ effectively compensates for everything
else that influences $T_{\mathrm{eff}}$.

Our mixing-length parameters also differ significantly from the spectroscopical
findings by \citet{Fuhrmann:1993p15161}, who concluded that one would
need an $\amlt$ with very low values with $\sim0.5$, to properly
fit hydrogen lines for various stars with the resulting temperature
stratifications. This, however, can be explained completely by the
fact that here only the outermost convective layers are traced, which
are not tested with our method for inferring $\amlt$ from the adiabatic
structure at the bottom of the convection zone, and that the mixing-length
parameter is indeed depth-dependent (see Sect.~\ref{sub:Depth-dependent-mml}).
This was already verified by \citet{Schlattl:1997p1676}.

\section{Mass mixing-length parameter\label{sec:Mass-mixing-length}}

\subsection{Deriving the mass mixing-length parameter\label{sub:Deriving-mml}}

In the following, we denote the temporal and spatial averaged thermodynamic
quantities with $\left\langle \dots\right\rangle $, which depict
only the $z$-dependence. Then, the momentum equation for a stationary
system yields 
\begin{eqnarray*}
\partial_{z}\left(\left\langle p_{\mathrm{th}}\right\rangle +\big<\rho v_{z}^{2}\big>\right) & = & \left\langle \rho\right\rangle g,
\end{eqnarray*}
where the divergence of the viscosity stress tensor vanishes on average.
This equation states that a given mass stratification ($\rho g$)
has to be supported by the joint thermodynamic ($p_{\mathrm{th}}$)
and turbulent pressure ($p_{\mathrm{turb}}=\rho v_{z}^{2}$) forces
to sustain equilibrium. Since the vertical velocity, $v_{z}$, appears
here, we solve for the latter and obtain 
\begin{eqnarray*}
\left\langle v_{z}\right\rangle  & \simeq & \sqrt{\frac{g-\left\langle p_{\mathrm{th}}\right\rangle /\left\langle \rho\right\rangle \partial_{z}\ln\left\langle p_{\mathrm{th}}\right\rangle }{\partial_{z}\ln\left\langle \rho\right\rangle +2\partial_{z}\ln\left\langle v_{z}\right\rangle }}.
\end{eqnarray*}
Here, we assume that $\big<\rho v_{z}^{2}\big>=\left\langle \rho\right\rangle \left\langle v_{z}\right\rangle ^{2}$,
but, we validated this equation with comparisons of averaged models.
Then, similar to the temperature gradient, $\nab=d\ln\left\langle T\right\rangle /d\ln\left\langle p_{\mathrm{tot}}\right\rangle $,
we introduce the notation for the gradient for a value $X$, but,
instead of the total pressure it is scaled by the thermodynamic pressure
scale height,

\begin{eqnarray*}
\nabla_{X} & = & \partial_{z}\ln\left\langle X\right\rangle /\partial_{z}\ln\left\langle p_{\mathrm{th}}\right\rangle ,
\end{eqnarray*}
and we can rewrite the vertical velocity to 
\begin{eqnarray}
\left\langle v_{z}\right\rangle  & \simeq & \sqrt{\frac{g/\partial_{z}\ln\left\langle p_{\mathrm{th}}\right\rangle -\left\langle p_{\mathrm{th}}\right\rangle /\left\langle \rho\right\rangle }{\nabla_{\rho}+2\nabla_{v_{z}}}}.\label{eq:vertical_velocity}
\end{eqnarray}
This equation depicts the correlation of the vertical velocity with
the gravity and pressure stratification, as well as the gradient of
the density and the gradient of the vertical velocity itself in the
hydrodynamic equilibrium. Now, we consider the gradient of the absolute
vertical mass flux, $\left\langle j_{z}\right\rangle =\left\langle \rho v_{z}\right\rangle $,
for the up- or downflows (because of conservation of mass, the mass
flux of the upflows, $j_{z}^{\uparrow}$, equals the mass flux of
the downflows, $j_{z}^{\downarrow}$) with

\begin{eqnarray*}
\nabla_{j_{z}} & = & \frac{\partial_{z}\ln|\big<j_{z}^{\mathrm{\uparrow\downarrow}}\big>|}{\partial_{z}\ln\left\langle p_{\mathrm{th}}\right\rangle },
\end{eqnarray*}
which indicates the length over which the up- or downflow has changed
by the $e$-fold, where the length scale is expressed in pressure
scale heights. \citet{Trampedach:2011p5920} introduced the mass mixing-length
as the inverse vertical mass flux scale height, that is $l_{m}=\partial_{z}\ln|\big<j_{z}^{\mathrm{\uparrow\downarrow}}\big>|^{-1}$,
which is in concordance with the gradient of the vertical mass flux
with $l_{m}=H_{P}/\nabla_{j_{z}}$. Furthermore, we define the mass
mixing-length parameter as the inverse gradient of the vertical mass
flux, 
\begin{eqnarray}
\am & \equiv & \nabla_{j_{z}}^{-1},\label{eq:alpha_m_ts11}
\end{eqnarray}
and we can decompose the gradient of the vertical mass flux into its
components and find 
\begin{eqnarray}
\am & \simeq & \left(\nabla_{\rho}+\nabla_{v_{z}}\right)^{-1},\label{eq:mass_mixing_length}
\end{eqnarray}
which states that the mass mixing-length parameter is the inverse
sum of the changes in the density and vertical velocity gradients.
The gradient of the filling factor also contributes, but, since it
vanishes in the deeper adiabatic convection zone and contributes only
very little confined to the photospheric transition region, we nelgect
this in our discussions \citep[see][]{Trampedach:2011p5920}. We note
that the definition in Eq. \ref{eq:alpha_m_ts11} is the same as introduced
by \citet{Trampedach:2011p5920}. Finally, we can now identify the
mass mixing-length parameter in the denominator of the vertical velocity
(Eq. \ref{eq:vertical_velocity}) and obtain the following expression:
\begin{eqnarray}
\left\langle v_{z}\right\rangle  & \simeq & \sqrt{\frac{\am}{1+\am\nabla_{v_{z}}}\left(\frac{g}{\partial_{z}\ln\left\langle p_{\mathrm{th}}\right\rangle }-\frac{\left\langle p_{\mathrm{th}}\right\rangle }{\left\langle \rho\right\rangle }\right)}.\label{eq:final_vertical_velocity}
\end{eqnarray}
This illustrates why the vertical velocity depends on the mass mixing-length
parameter, similarly to the MLT velocity $v_{\mathrm{MLT}}$, which
depends on mixing-length parameter with $v_{\mathrm{MLT}}\propto\amlt$
(see Eq. \ref{eq:convective velocity}).

To complete the comparison of the mass mixing-length parameter with
the (MLT) mixing-length parameter, we derive its dependence on the
convective energy flux. We assume that the mean convective energy
flux consists of the fluctuations of the total energy ($\varepsilon_{\mathrm{tot}}=\varepsilon+p_{\mathrm{th}}/\rho+\vec{v}^{2}/2$),
which we depicted with $f$, and is carried by the mean vertical mass
flux, that is 
\begin{eqnarray*}
\left\langle F_{\mathrm{conv}}\right\rangle  & \sim & \left\langle f\right\rangle \left\langle \rho v_{z}\right\rangle ,
\end{eqnarray*}
where we assume that $v_{z}$ is the hydrodynamic velocity given in
Eq. \ref{eq:vertical_velocity} and also that $\big<\rho v_{z}\big>=\left\langle \rho\right\rangle \left\langle v_{z}\right\rangle $.
We determine the divergence of the convective energy flux, $\partial_{z}\left\langle F_{\mathrm{conv}}\right\rangle $,
and solve for the total energy fluctuations, which yields 
\begin{eqnarray*}
f & \simeq & \frac{1}{\nabla_{\rho}+\nabla_{v_{z}}}\frac{\partial_{z}\left\langle F_{\mathrm{conv}}\right\rangle /\left\langle \rho v_{z}\right\rangle +\partial_{z}\left\langle f\right\rangle }{\partial_{z}\ln\left\langle p_{\mathrm{th}}\right\rangle }.
\end{eqnarray*}
Then, we can substitute the convective energy losses, $\partial_{z}\left\langle F_{\mathrm{conv}}\right\rangle $,
by the radiative $ $cooling rate, $-\left\langle q_{\mathrm{rad}}\right\rangle $,
because of conservation of total energy, and we can identify the mass
mixing-length parameter in the convective energy flux as well and
obtain 
\begin{eqnarray}
\left\langle F_{\mathrm{conv}}\right\rangle  & \simeq & -\am\frac{\left(\left\langle q_{\mathrm{rad}}\right\rangle +\left\langle \rho v_{z}\right\rangle \partial_{z}\left\langle f\right\rangle \right)}{\partial_{z}\ln\left\langle p_{\mathrm{th}}\right\rangle }.\label{eq:final_fconv}
\end{eqnarray}
This equation is basically the expression for the conservation of
energy. These two equations for the velocity and the convective energy
flux are just reformulated approximations of the hydrodynamic mean-field
equations. To close this set of equations, one still would need information
about the gradient of the velocity and total energy fluctuation, as
well as the radiative cooling rates.

\subsection{Depth-dependence of the mass mixing-length parameter\label{sub:Depth-dependent-mml}}

\begin{figure}
\includegraphics[width=88mm]{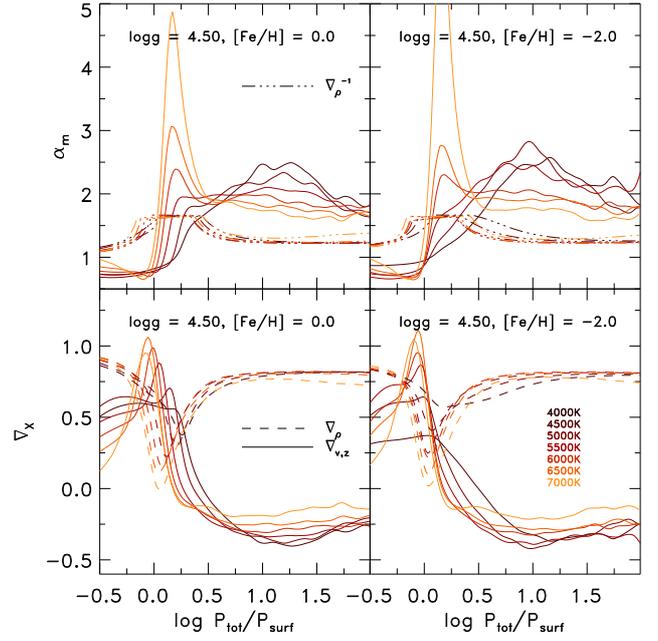}

\caption{\label{fig:mml_comp}\emph{Top panel}: The mass mixing-length parameter
$\am$ from Eq. \ref{eq:mass_mixing_length} (solid) and the inverse
gradient of the density, $\nabla_{\rho}^{-1}$ (triple-dotted dashed
line). For clarity we excluded values with $\nabla_{\rho}^{-1}>5/3$
just below the optical surface ($0<\log p_{\mathrm{tot}}/p_{\mathrm{surf}}<0.5$).
\emph{Bottom panel}: the gradient for density, $\nabla_{\rho}$, and
vertical velocity, $\nabla_{v_{z}}$, (dashed and solid lines, respectively)
for different stellar parameters.}
\end{figure}
\begin{figure*}
\includegraphics[width=88mm]{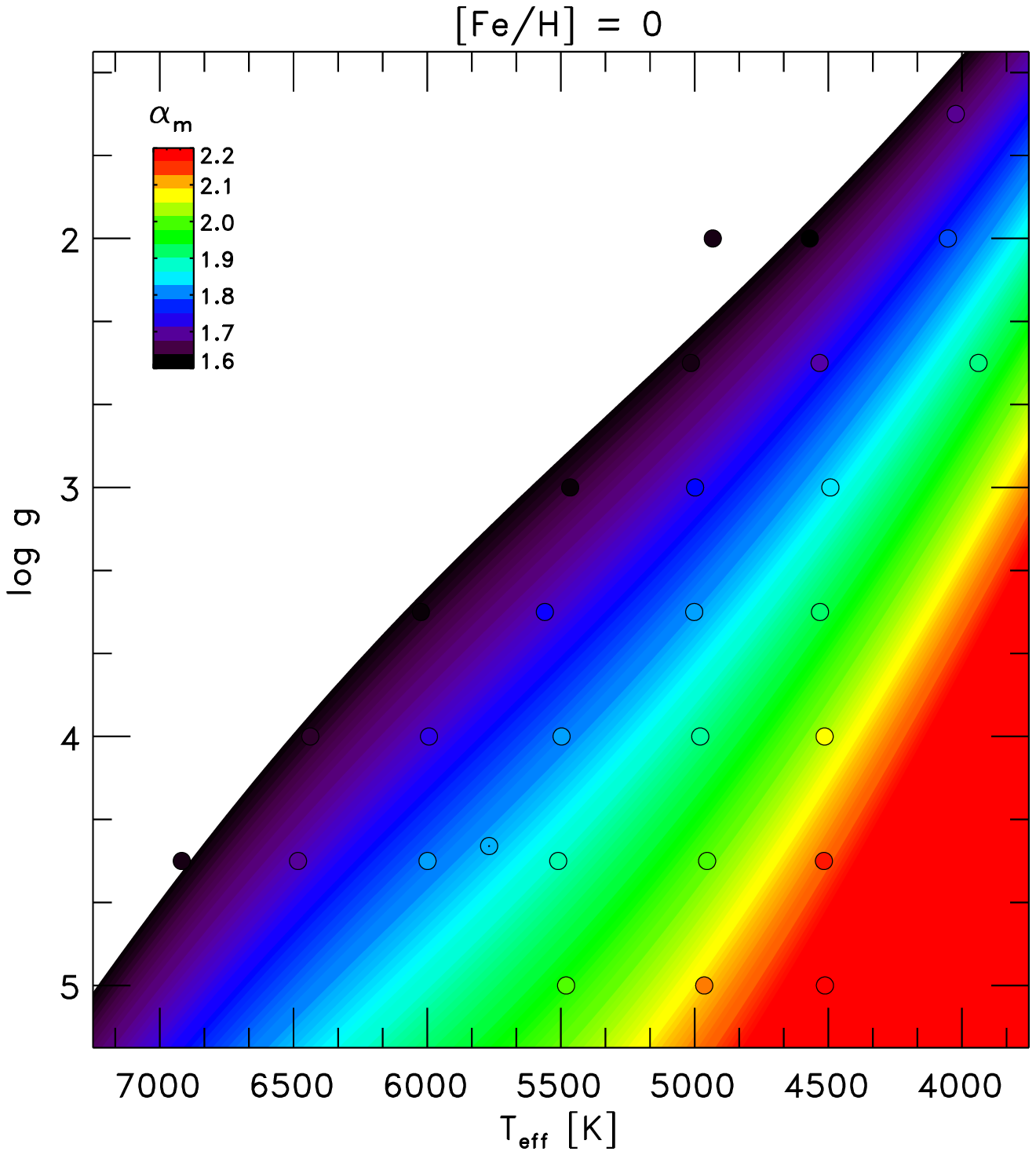}\includegraphics[width=88mm]{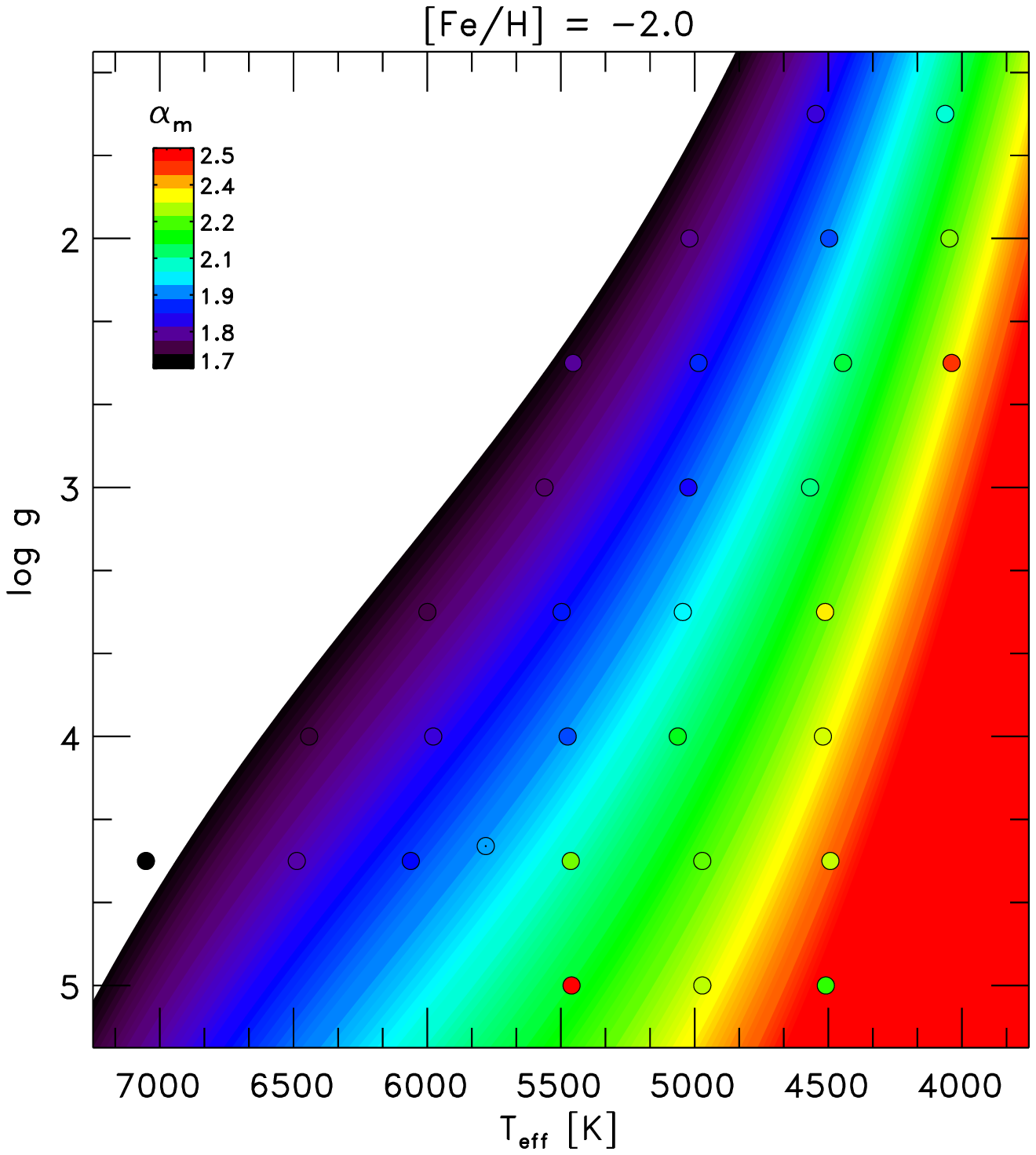}

\caption{\label{fig:mml_ov}As Fig. \ref{fig:alpha_mlt_atmo}, but here the
mass mixing-length parameter $\am$ is shown.}
\end{figure*}
\begin{figure*}
\includegraphics[width=176mm]{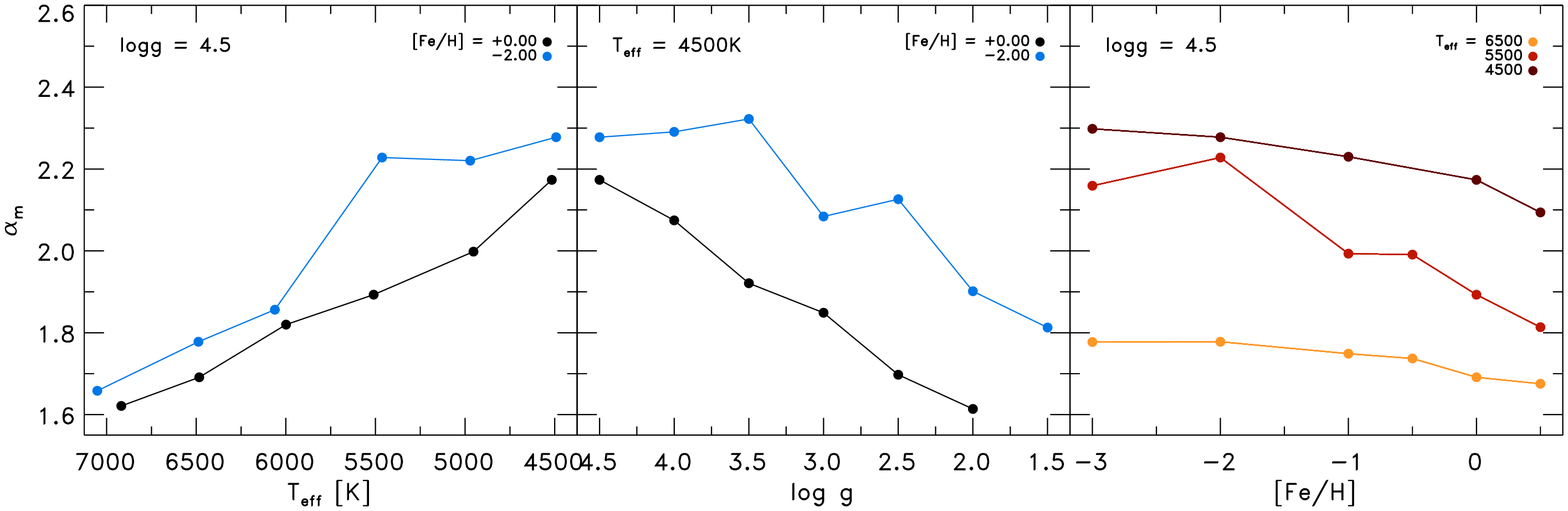}

\caption{\label{fig:mmlr_stellar_parameter}As Fig. \ref{fig:alpha_mlt_stellar_parameter},
but here the mass mixing-length parameter $\am$ is shown.}
\end{figure*}
\begin{figure}
\includegraphics[width=88mm]{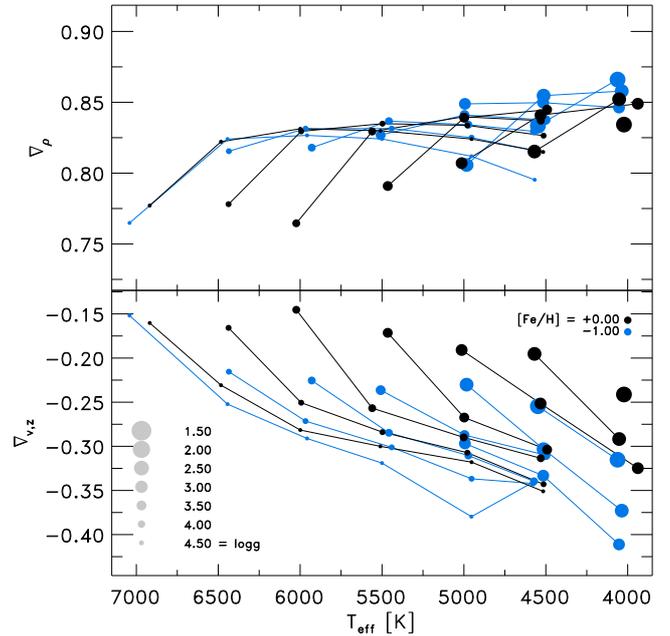}

\caption{\label{fig:ov_mml_comp}Mean gradient of the density, $\nabla_{\rho}$,
and the vertical velocity, $\nabla_{v_{z}}$, in the convection zone
for different stellar parameters.}
\end{figure}
Following \citet{Trampedach:2011p5920}, we tried to derive the mass
mixing-length parameter from the vertical mass flux of the downflows
(Eq. \ref{eq:alpha_m_ts11}), but, we found that the fluctuations
in the vertical velocity field are enhanced in our simulations, which
is probably caused by the higher numerical vertical resolution (the
simulations by \citealt{Trampedach:2013ApJ...769...18T} have a thrice
lower vertical resolution and therefore exhibit fewer turbulent and
more laminar structures in the downflows). We found the rms vertical
velocity to be less sensitive to the statistical fluctuations in the
deeper convection zone, therefore, we derived the mass mixing-length
by using the gradient of the rms vertical velocity in Eq. \ref{eq:mass_mixing_length},
instead of deriving the mass mixing-length from the vertical mass
flux of the downflows (Eq. \ref{eq:alpha_m_ts11}). Therefore, a comparison
with \citet{Trampedach:2011p5920} is only qualitatively meaningful.

In Fig.~\ref{fig:mml_comp}, we illustrate the horizontally and temporally
averaged, depth-dependent mass mixing-length parameter for different
stellar parameters, which we derived from our 3D RHD simulations.
In the convection zone, the mass mixing-length parameter has values
around $\sim2$, while above the optical surface, $\am$ has lower
values around $\sim0.5$. \citet{Fuhrmann:1993p15161} found that
similar low values for the mixing-length parameter $\amlt$ yield
better fits for Balmer lines, but, they also used high values for
the temperature distribution parameter with $y=0.5$ (see also App.
\ref{app:Additional-MLT-parameters}), and moreover, the influence
of $\amlt$ becomes negligible towards the optical surface, where
the Balmer lines form (see Fig. \ref{fig:match_alpha_mlt_atmo}).
Therefore, the agreement of the depth-dependent $\am$ with their
low values for $\amlt$ might be just a coincidence. Furthermore,
just below the optical surface ($\log p_{\mathrm{tot}}/p_{\mathrm{surf}}=0$)
at the photospheric transition region, $\am$ features a peak, which
depends on the stellar parameters, in particular, for higher $\teff$,
the peak in between increases, while in the convection zone it is
the flatter. We remark that the peak in $\am$ coincides with the
location of the peak in the $\vzrms$. We also included the inverse
gradient of density in the same figure with $\am$, demonstrating
that the adiabatic value of $\am$ in the convection zone is mainly
contributed by the density gradient.

We also show the gradients of the density and vertical velocity in
Fig.~\ref{fig:mml_comp}, which are the two main components of $\am$.
The gradients of the filling factors also contribute to the variation
of mass mixing-length. However, similar to the findings by \citet{Trampedach:2011p5920},
we find that the fillings factors are constant in the convection zone,
therefore, their contribution is negligible. The variation of $\am$
in the convection zone arises mainly because of the different velocity
gradients, since the density gradient converges always to very similar
adiabatic values ($\gamma_{\mathrm{ad}}\simeq\nabla_{\rho}^{-1}$).
For a monoatomic ideal gas with radiation pressure the adiabatic exponent
is given by $\gamma_{\mathrm{ad}}=\left(1-\nabla_{\mathrm{ad}}\right)^{-1}$,
and with $\nabla_{\mathrm{ad}}=1/4$ one obtains $\gamma_{\mathrm{ad}}\sim4/3$
\citep[see]{Kippenhahn:2013sse..book.....K}. $\nabla_{\rho}^{-1}$
it is close to 1.2 (see Fig.~\ref{fig:mml_comp}). For a nonideal
gas differences due to nonideal effects are to be expected. On the
other hand, $\nabla_{\rho}$ is close to $\sim0.8$ therefore, similar
to a value for an ideal gas with $3/4$, while $\nabla_{v_{z}}$ is
between $-0.4$ and $-0.15$ (see also Fig.~\ref{fig:ov_mml_comp}).

In the vicinity of the optical surface, the cooling rates are imprinted
on the gradients for the density and velocity with a sharp transition.
Towards the interior, the density increases because of the stratification
and hydrostatic equilibrium, hence the gradient is $\nabla_{\rho}>0$,
while the velocity decreases, and therefore $\nabla_{v_{z}}<0$. The
signs of $\nabla_{\rho}$ and $\nabla_{v_{z}}$ are opposite because
of the conservation of mass. In the interior, the stellar fluid is
compressed and the velocity slows down, meaning that the convective
energy is carried with a slower, thicker mass flux. For higher $\teff$,
the (negative) velocity gradient has a lower amplitude and is therefore
closer to zero, and a smaller amplitude of $\nabla_{v_{z}}$ implies
a steeper drop of the vertical velocity towards the interior, which
also entails a larger maximum of the vertical velocity (see Fig.~\ref{fig:ov_mml_comp}).
The velocity gradient reduces the density gradient, but, a lower sum
of $\nabla_{\rho}$ and $\nabla_{v_{z}}$ relates to a higher $\am$
because of the inverse relation (see Fig.~\ref{fig:mml_ov}). Since
the density gradient is very similar for different stellar parameters,
the variation in $\am$ arises mainly from the differences in the
velocity gradient. Therefore, we can relate the variation of the entropy
jump with the variation of the velocity gradient, that is $\Delta s\sim e^{\nabla_{v_{z}}}$,
which was also concluded by \citet{Trampedach:2011p5920} for the
mass mixing-length parameter in an extended solar simulation.

\subsection{Mean mass mixing-length parameter in the convection zone\label{sub:Mean-mass-mixing-length}}

We determined the \emph{mean} mass mixing-length parameter of the
convection zone below the optical surface between the location of
the peak in the density scale height, that is $\max\left(\partial_{z}\ln\rho\right)_{\tau>1}$,
and the bottom, but, avoided bottom boundary effects on the vertical
velocity. We performed linear fits of the density and vertical rms-velocity
gradients by considering all snapshots, and from both gradients we
determined the mean value of $\am$ as given in Eq. \ref{eq:mass_mixing_length}.
We note that our method of retrieving a mean value differs from that
by \citet{Trampedach:2011p5920}. The convection zones in the 3D simulations
have to be extended enough, so that lower boundary effects on the
vertical velocities are minimized, which is the case for most models,
except for some metal-poor giants that are slightly too shallow to
properly match $\am$.

The results for $\am$ are displayed in Fig.~\ref{fig:mml_ov}, while
in Fig.~\ref{fig:ov_mml_comp} we depict the mean values of the density
and velocity gradients. From the solar simulation we determined $\am^{\odot}=1.83$,
which is close to the solar mass mixing-length parameter by \citet{Trampedach:2011p5920}
with $1.76$. Furthermore, the mass mixing-length parameter depicts
qualitatively very similar systematic variations with stellar parameter,
as we found for $\amlt$ above. In particular, it decreases for higher
$\teff$ and $\feh$, and lower $\logg$, and the range in $\am$
between $\sim1.7$ and $\sim2.3$ is qualitatively similar to that
of $\amlt$ (see also Fig. \ref{fig:mmlr_stellar_parameter}). In
general, we find values for $\am$ that are qualitatively similar
to those found by \citet{Trampedach:2011p5920}, in particular, the
dwarf models ($\logg=4.5$) have a similar slope with $\teff$. As
mentioned above, we consider the mass mixing-length parameter from
the gradients of the density and rms vertical velocity (see Eq. \ref{eq:mass_mixing_length})
instead of the unidirectional mass flux, and we also used a different
method for determining a mean value, therefore, differences in the
results are to be expected.

\begin{figure}
\includegraphics[width=88mm]{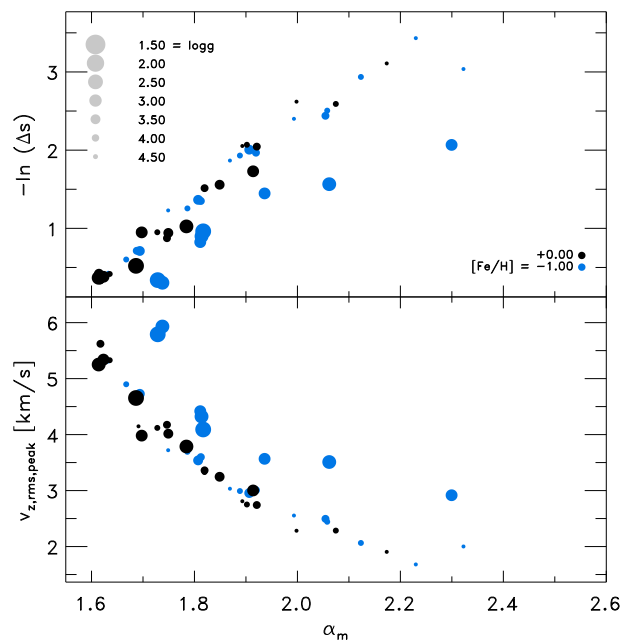}

\caption{\label{fig:mml_anticorrelation}Correlation of the mass mixing-length
parameter, $\am$, with the logarithmic inverse of the entropy jump
$-\ln\left(\Delta s\right)$ and the peak of the vertical rms-velocity
(top and bottom panel, respectively).}
\end{figure}
The variation of $\am$ is also similar to the logarithmic inverse
variation of the entropy jump. In Fig.~\ref{fig:mml_anticorrelation}
we compare $\am$ with $ $the logarithmic inverse entropy jump and
find a similar tight correlation between the two as we found for the
mixing-length parameter $\amlt\left(\Delta s\right)$ above (Sect.
\ref{sub:matching_ssj_atmo}). The stronger deviations for the metal-poor
giants originate from the fact that these models are slightly shallower,
therefore, the match of the mass mixing-length parameter is perturbed
because of the lower boundary effects on the velocity. We also illustrate
the tight anticorrelation of the peak vertical rms-velocity with the
mass mixing-length parameter in Fig. \ref{fig:mml_anticorrelation}.

\begin{figure}
\includegraphics[width=88mm]{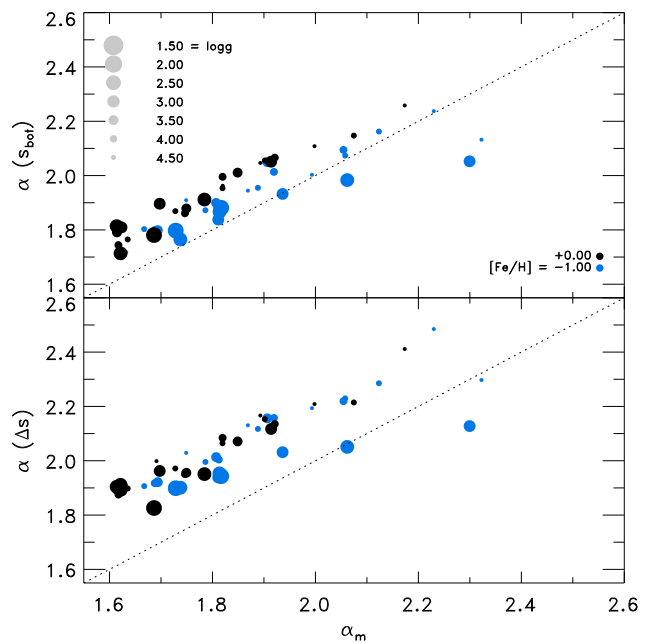}

\caption{\label{fig:comp_mml_alpha_ssj}Comparison of the mass mixing-length
parameter, $\am$, with the mixing-length parameter calibrated with
$\ssbot$ and $\Delta s$ (top and bottom panel, respectively).}
\end{figure}
A comparison of the mass mixing-length parameter with the mixing-length
parameter calibrated with the entropy of the deep adiabatic convection
zone and the entropy jump is shown in Fig.~\ref{fig:comp_mml_alpha_ssj},
and these also correlate well. The mixing-length parameters are slightly
higher than $\am$ with a systematic offset around $\sim0.1$ and
$\sim0.2$, which smaller for $\amlt\left(\ssbot\right)$ than for
$\amlt\left(\Delta s\right)$. This illustrates that the mixing-length
parameter in the framework of MLT has a physical background that originates
in the mass mixing-length parameter (or inverse vertical mass flux
gradient). However, since the MLT is incomplete, a one-to-one correspondence
between $\amlt$ and $\am$ is hardly expected; nonetheless, the good
agreement between the two is an interesting result.

\section{Velocity correlation length\label{sec:velocity_correlation_length}}

The physical interpretation of the mixing-length parameter is conceptually
the mean free path of a convective eddy over which it can preserve
its identity before it resolves into its environment. In a real stratified
hydrodynamic fluid the spatial two-point (auto)correlation function
of the vertical velocity can be regarded as the 3D analog of the mixing-length
parameter $\amlt$ as proposed by \citet{Chan:1987p22755}. The two-point
correlation function for the values $q_{1}$ and $q_{2}$ is given
by 
\begin{eqnarray}
C\left[q_{1},q_{2}\right] & = & \frac{\left\langle q_{1}q_{2}\right\rangle -\left\langle q_{1}\right\rangle \left\langle q_{2}\right\rangle }{\sigma_{1}\sigma_{2}},\label{eq:correlation_function}
\end{eqnarray}
with $\sigma_{i}$ being the the standard deviation of $q_{i}$ and
$\left\langle \dots\right\rangle $ is the spatial horizontal average.
To derive the vertical correlation function of the convective velocity
field, we considered the vertical component of the velocity field,
$v_{z}$, of a single fixed layer $z_{0}$ and derived the correlation
functions for all other layers $z_{i}$, i.e. $C\left[v_{z_{0}},v_{z_{i}}\right]$,
which was performed for twenty equidistant layers and covered the
whole vertical depth scale of the simulation box.

\begin{figure}
\includegraphics[width=88mm]{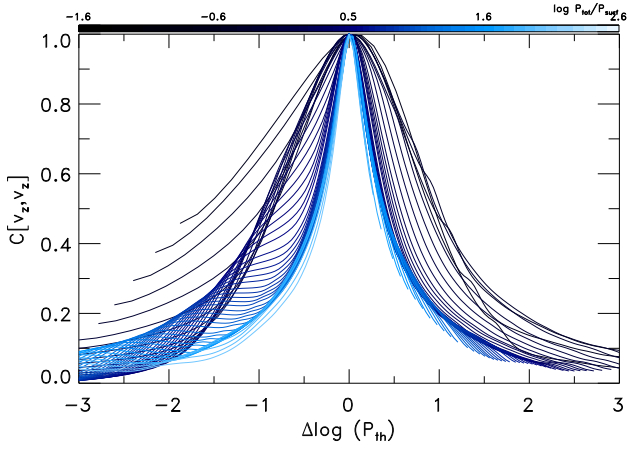}

\caption{\label{fig:vcf_sun1}Vertical two-point correlation function of the
vertical velocity, $C\left[v_{z},v_{z}\right]$, vs. the difference
in the thermodynamic pressure, $\Delta\log P_{\mathrm{th}}$, for
the solar simulation. The different heights are indicated with a blue
color-coding. Note the convergence of the correlation width in the
convection zone against an adiabatic value.}
\end{figure}

\begin{figure}
\includegraphics[width=88mm]{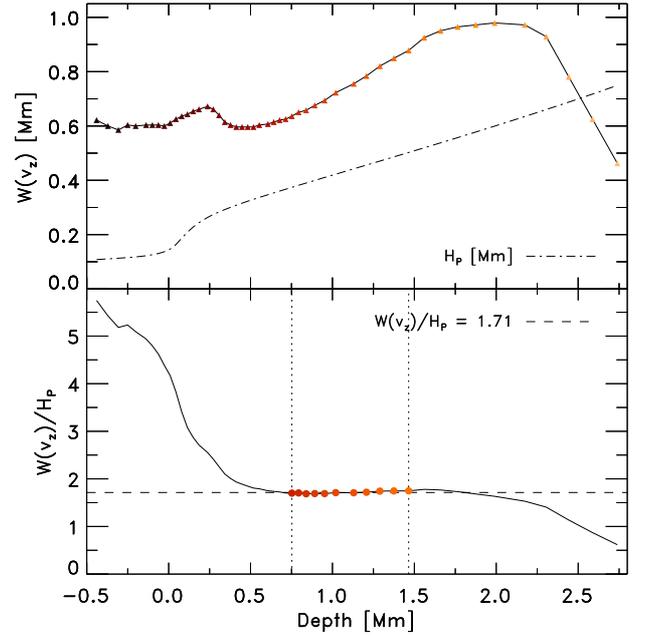}\caption{\label{fig:vcf_sun2}\emph{Top panel}: Vertical correlation length
of the vertical velocity (solid line with triangles) and pressure
scale height (dotted-dashed line) shown against the depth for the
solar model. We indicate the different heights with the same color-coding
as used in Fig.~\ref{fig:vcf_sun1}. \emph{Bottom panel}: Vertical
correlation length scaled by the pressure scale height, which yields
an average of $1.71$ (dashed line) in the considered region for averaging
the correlation length (vertical dotted lines with filled circles)
for the solar simulation.}
\end{figure}
In Fig.~\ref{fig:vcf_sun1} we show the two-point correlation function
of the vertical velocity field, $C\left[v_{z},v_{z}\right]$, derived
for the solar simulation for the individual snapshots and then temporally
averaged. For convenience, the correlation function is shown in differences
of logarithmic pressure to the considered layer, $\Delta\log P_{\mathrm{th}}=\log P_{\mathrm{th}}(z_{0})-\log P_{\mathrm{th}}(z_{i})$.
Then, the correlation function always reaches unity for $z_{i}=z_{0}$
and has a Gaussian-like shape. Furthermore, it is broader above the
optical surface ($p_{\mathrm{tot}}/p_{\mathrm{surf}}=1$), which is
due to the rapid decline of the pressure scale height; while below
the latter the width seems to converge on a certain adiabatic value
(see Fig.~\ref{fig:vcf_sun2}). When one considers the width of the
correlation function in geometrical depth, instead of pressure, then
$W\left(v_{z}\right)$ is constant around $\sim0.6\,\mathrm{Mm}$
from the top down to $\sim0.5\,\mathrm{Mm}$ and increases then with
a fixed multiple ($1.71$) of the pressure scale height (see Fig.
\ref{fig:vcf_sun2}), which is the same as \citet{Robinson:2003p17201}
found. The higher values for $W\left(v_{z}\right)/H_{P}$ above $0.5\,\mathrm{Mm}$
result from the lower $H_{P}$.

\begin{figure}
\includegraphics[width=88mm]{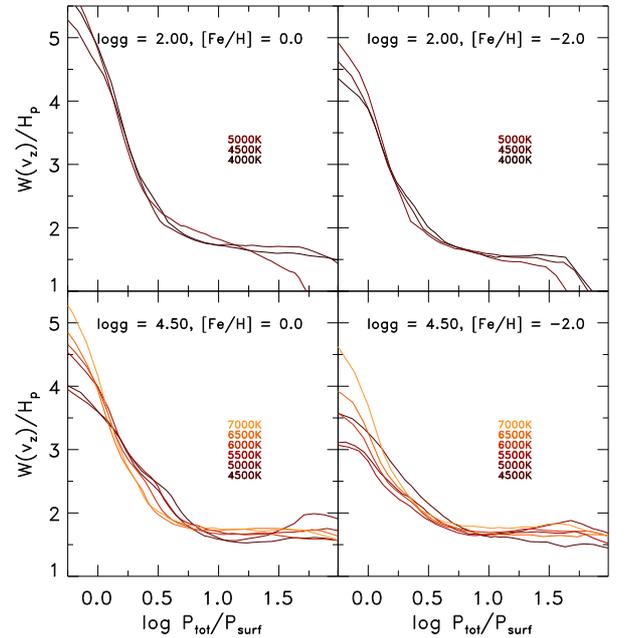}\caption{\label{fig:ov_wuy}Correlation length of the vertical velocity vs.
surface normalized pressure for different stellar parameters.}
\end{figure}
The full-width at half maximum (FWHM) of the two-point correlation
function of the vertical velocity, $C\left[v_{z},v_{z}\right]$, which
we denote with $W\left(v_{z}\right)$, gives an estimate on the size
or length scale of the coherent vertical structures. The characteristic
local length scale for the turbulent convective eddies can be determined
with $W\left(v_{z}\right)$. With the term vertical correlation length
we refer to $W\left(v_{z}\right)$. Similar to the mixing-length,
it is preferable to scale the correlation length by the pressure scale
height, that is $W\left(v_{z}\right)/H_{P}$, since the the latter
increases towards deeper layers. Then, for the solar simulation (see
Fig.~\ref{fig:vcf_sun2}) the converging value for the width is $W\left(v_{z}\right)/H_{P}=1.71$.
This means that the coherent vertical structures extend $1.71H_{P}$
in the convection zone, and this value is similar to the mixing-length
parameter ($\amlt=1.94$). \citet{Chan:1987p22755,Chan1989ApJ...336.1022C}
also found a similar scaling of $C\left[v_{z},v_{z}\right]$ with
pressure scale height in a 3D simulation for the Sun. For different
stellar parameters we find a rather similar convergence of the correlation
length of the vertical velocity in the convection zone (see Fig.~\ref{fig:ov_wuy}).
\citet{Ludwig:2006p7586} found similar values for the correlation
length of the vertical velocity with $W\left(v_{z}\right)/H_{P}\sim2$
in the vicinity of the lower boundary for a number of different simulations,
while \citet{Viallet:2013ApJ...769....1V} recently found for a red-giant
simulation that the vertical correlation length of the vertical velocity
scales with approximately twice of the pressure scale height.

\begin{figure}
\includegraphics[width=88mm]{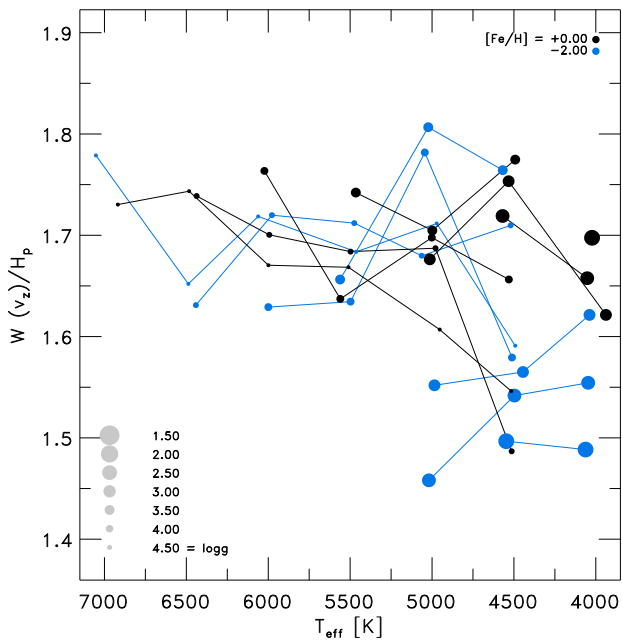}\caption{\label{fig:psg_wuy}Overview of mean vertical correlation length of
the vertical velocity in the convection zone for different stellar
parameters.}
\end{figure}
We also determined the mean value of the correlation length in the
convection zone below $\log p_{\mathrm{tot}}/\log p_{\mathrm{surf}}>1$.
Close to the bottom boundary, the correlation function will increasingly
overturn because we lack information in the deeper layers. Therefore,
we applied for a mean correlation length a cut at the bottom, where
$W\left(v_{z}\right)/H_{P}$ begins to decrease (see Fig.~\ref{fig:vcf_sun2}).

The resulting mean values of $W\left(v_{z}\right)/H_{P}$ for different
stellar parameters are depicted in Fig.~\ref{fig:psg_wuy}. They
are distributed between $\sim1.5$ and $\sim1.8$. This is an interesting
result, since it confirms, to a certain extent, the physical motivation
for the mixing-length parameter, $\amlt$: the vertical velocity field,
hence the vertical mass flux, correlates similarly with the pressure
scale heights in the convection zone. However, the variation of $W\left(v_{z}\right)/H_{P}$
with stellar parameters (Fig.~\ref{fig:psg_wuy}) is not as clear
and systematical as we found above for $\amlt$ and $\am$ (see Sect.
\ref{sec:Mass-mixing-length}).

Furthermore, in contrast to the mixing-length parameter ($\amlt$
and $\am$), the correlation length seems to increase for higher $\teff$.
The reason for this might be the horizontal granule size, which we
found to decrease slightly for lower effective temperatures, since
the pressure scale height decreases (see Paper I). Moreover, the granular
cells, which can be highlighted with the temperature excess from the
background, feature distinct regular flat cylindric or pillar-like
topologies.

Finally, we considered the correlation length of other variables and
found that the horizontal velocity is rather similar, but with slightly
lower correlation length with $\sim1.4$. In addition, the entropy,
temperature, and pressure have values around $\sim1.3$, while the
value for the density is close to unity.

\section{Conclusions\label{sec:Conclusions}}

We have calibrated the mixing-length parameter using realistic 3D
RHD simulations of stellar surface convection by employing a 1D MLT
stellar atmosphere code with identical microphysics. The calibration
was achieved by varying the mixing-length parameter and matching the
adiabatic entropy value of the deeper convection zone, $\ssbot$,
or alternatively, matching the entropy jump, $\Delta s$. In both
ways we found the mixing-length to decrease for higher $\teff$ and
$\feh$, and lower $\logg$. The mixing-length varies in the range
of $1.7-2.3$ for $\amlt\left(\ssbot\right)$ and $\sim1.8-2.4$ for
$\amlt\left(\Delta s\right)$, and will lead to differences of up
to $\pm20\,\%$ in $\amlt$ depending on the stellar mass. This changes
the stellar interior structure by extending or shortening the depth
of the convection zone and thus the stellar evolution; we intend to
investigate in future studies how in detail a realistic $\amlt$ will
impact basic stellar evolution predictions.

Furthermore, we derived from the hydrodynamic mean field equations
(for the first-time) a physically motivated connection of the mass
mixing-length, which is the inverse of the vertical mass flux gradient,
with the mixing-length. We determined the mass mixing-length parameter
and found that it varies qualitatively similar to the mixing-length
parameter in the range of $1.6-2.3$. The mass mixing-length parameter
is also depth-dependent and decreases above the surface to lower values
around $\sim0.5$, which agrees with previous findings from observations.
Finally, the mass mixing-length parameter and mixing-length parameter
strongly correlate with the logarithmic inverse of the entropy jump
for different stellar parameters, that is $\amlt\sim-\ln\Delta s$.
Finally, we also derived the vertical velocity correlation length,
which features values similar to that of the mixing-length with approximately
$\sim1.6-1.8$ of pressure scale height, but, the dependence with
$\teff$ is inverted, meaning that the correlation length decreases
with $\teff$.

To summarize the importance of our work: we can finally remove the
free parameters inherent in MLT and also avoid having to use solar
calibrations for other stars.
\begin{acknowledgements}
We thank Regner Trampedach, {\AA}ke Nordlund, and Bob Stein for helpful
discussions. We acknowledge access to computing facilities at the
Rechenzentrum Garching (RZG) of the Max Planck Society and at the
Australian National Computational Infrastructure (NCI), where the
3D RHD simulations were carried out. 
\end{acknowledgements}
\bibliographystyle{aa}
\bibliography{papers}

\appendix

\section{Tables\label{app:Tables}}

In Table \ref{tab:table_mixing_length}, we list results for the solar
metallicity. The complete table is available at CDS \href{http://cdsarc.u-strasbg.fr}{cdsarc.u-strasbg.fr}
and also at \href{http://www.stagger-stars.net}{www.stagger-stars.net}.
\begin{table*}
\caption{\label{tab:table_mixing_length}Stellar parameters: effective temperature,
$\teff$, and surface gravity, $\logg$ (Cols. 1 and 2 in $\left[\mathrm{K}\right]$
and $\left[\mathrm{dex}\right]$).}

\begin{tabular}{llllllllllllll}
\hline\hline
$T_{\rm{eff}}$ & $\log g$ & $\lg \rho_{\mathrm{bot}}$ & $\lg T_{\mathrm{bot}}$ &  $\lg p_{\mathrm{th}}^{\mathrm{bot}}$ & $\ssbot$ & $\Delta s$ & $\delta s_{\mathrm{rms}}^{\mathrm{peak}}$ & $\delta \rho_{\mathrm{rms}}^{\mathrm{peak}}$ & $\delta v_{z,\mathrm{rms}}^{\mathrm{peak}}$ &$\amlt^{\ssbot}$ & $\amlt^{\Delta s}$ & $\am$ & $W(v_z)/H_P$ \\
\hline
      4023 &    1.50 &      0.717 &      4.272 &      1.061 &      2.304 &      0.594 &     14.023 &     46.897 &      0.465 &      1.781 &      1.826 &      1.686 &      1.698\\
      4052 &    2.00 &      1.125 &      4.233 &      1.368 &      2.018 &      0.359 &      9.088 &     35.288 &      0.379 &      1.912 &      1.951 &      1.784 &      1.658\\
      3938 &    2.50 &      1.691 &      4.239 &      1.889 &      1.776 &      0.177 &      4.733 &     24.764 &      0.300 &      2.051 &      2.117 &      1.914 &      1.621\\
      4569 &    2.00 &      0.679 &      4.342 &      1.120 &      2.417 &      0.692 &     16.099 &     50.583 &      0.525 &      1.814 &      1.904 &      1.614 &      1.719\\
      4532 &    2.50 &      1.357 &      4.279 &      1.669 &      2.039 &      0.387 &      9.935 &     36.301 &      0.398 &      1.896 &      1.962 &      1.697 &      1.753\\
      4492 &    3.00 &      1.785 &      4.266 &      2.029 &      1.808 &      0.211 &      5.783 &     26.808 &      0.325 &      2.011 &      2.071 &      1.849 &      1.775\\
      4530 &    3.50 &      2.103 &      4.269 &      2.322 &      1.682 &      0.129 &      3.707 &     20.261 &      0.274 &      2.066 &      2.135 &      1.921 &      1.656\\
      4513 &    4.00 &      2.419 &      4.277 &      2.625 &      1.580 &      0.075 &      2.205 &     14.454 &      0.229 &      2.147 &      2.215 &      2.075 &      1.487\\
      4516 &    4.50 &      2.721 &      4.292 &      2.927 &      1.503 &      0.045 &      1.348 &      9.928 &      0.191 &      2.258 &      2.411 &      2.173 &      1.546\\
      4512 &    5.00 &      3.013 &      4.308 &      3.226 &      1.436 &      0.029 &      0.830 &      6.377 &      0.154 &      2.342 &      2.907 &      2.221 &      1.684\\
      4932 &    2.00 &      0.042 &      4.535 &      0.700 &      2.766 &      0.948 &     21.258 &     69.923 &      0.725 &      1.714 &      1.911 &      1.621 &      1.630\\
      5013 &    2.50 &      0.883 &      4.374 &      1.358 &      2.381 &      0.683 &     16.250 &     51.278 &      0.534 &      1.809 &      1.890 &      1.623 &      1.676\\
      4998 &    3.00 &      1.534 &      4.308 &      1.882 &      2.026 &      0.390 &     10.116 &     35.942 &      0.402 &      1.879 &      1.956 &      1.749 &      1.705\\
      5001 &    3.50 &      1.960 &      4.295 &      2.243 &      1.805 &      0.220 &      6.134 &     27.035 &      0.336 &      1.995 &      2.084 &      1.820 &      1.698\\
      4978 &    4.00 &      2.292 &      4.293 &      2.538 &      1.662 &      0.126 &      3.700 &     19.745 &      0.275 &      2.056 &      2.153 &      1.902 &      1.687\\
      4953 &    4.50 &      2.604 &      4.301 &      2.837 &      1.561 &      0.073 &      2.196 &     14.015 &      0.228 &      2.108 &      2.209 &      1.998 &      1.607\\
      4963 &    5.00 &      2.885 &      4.314 &      3.118 &      1.487 &      0.045 &      1.398 &      9.888 &      0.185 &      2.143 &      2.227 &      2.130 &      1.577\\
      5465 &    3.00 &      1.084 &      4.403 &      1.589 &      2.343 &      0.657 &     15.917 &     48.856 &      0.527 &      1.791 &      1.893 &      1.614 &      1.742\\
      5560 &    3.50 &      1.663 &      4.345 &      2.062 &      2.043 &      0.417 &     10.870 &     37.436 &      0.418 &      1.861 &      1.951 &      1.747 &      1.637\\
      5497 &    4.00 &      2.139 &      4.322 &      2.456 &      1.791 &      0.221 &      6.244 &     26.379 &      0.333 &      1.953 &      2.065 &      1.820 &      1.684\\
      5510 &    4.50 &      2.486 &      4.322 &      2.769 &      1.649 &      0.128 &      3.791 &     19.527 &      0.281 &      2.047 &      2.167 &      1.893 &      1.668\\
      5480 &    5.00 &      2.791 &      4.330 &      3.060 &      1.548 &      0.076 &      2.343 &     14.272 &      0.225 &      2.068 &      2.186 &      2.002 &      1.670\\
      5768 &    4.44 &      2.367 &      4.336 &      2.688 &      1.725 &      0.184 &      5.313 &     23.788 &      0.308 &      1.979 &      2.089 &      1.825 &      1.702\\
      6023 &    3.50 &      1.130 &      4.493 &      1.737 &      2.403 &      0.715 &     17.022 &     51.883 &      0.562 &      1.744 &      1.875 &      1.617 &      1.764\\
      5993 &    4.00 &      1.865 &      4.364 &      2.281 &      1.994 &      0.387 &     10.302 &     35.917 &      0.412 &      1.869 &      1.971 &      1.728 &      1.700\\
      5998 &    4.50 &      2.301 &      4.344 &      2.644 &      1.771 &      0.218 &      6.225 &     25.994 &      0.332 &      1.962 &      2.081 &      1.820 &      1.670\\
      6437 &    4.00 &      1.384 &      4.495 &      1.989 &      2.322 &      0.659 &     16.117 &     48.227 &      0.533 &      1.765 &      1.898 &      1.635 &      1.739\\
      6483 &    4.50 &      2.008 &      4.386 &      2.448 &      1.972 &      0.378 &     10.054 &     34.007 &      0.415 &      1.889 &      1.998 &      1.691 &      1.744\\
      6918 &    4.50 &      1.545 &      4.543 &      2.201 &      2.301 &      0.652 &     15.738 &     46.039 &      0.526 &      1.796 &      1.911 &      1.621 &      1.730\\
\hline
\end{tabular}

Notes: The conditions at the lower boundary: density, $\rho$, temperature,
$T$, pressure, $p_{\mathrm{th}}$, entropy at the bottom; the entropy
jump, $\Delta s$; the peak fluctuations in: entropy, $\delta s_{\mathrm{rms}}^{\mathrm{peak}}$,
density, $\delta\rho_{\mathrm{rms}}^{\mathrm{peak}}$, vertical velocity
$\vzrmsp$; the mixing-length: $\amlt\left(\ssbot\right)$ and $\amlt\left(\Delta s\right)$;
mass mixing-length parameter, $\am$, and correlation length $W\left(v_{z}\right)/H_{P}$. 
\end{table*}

\section{Functional fits\label{app:Functional-fits}}

Similar to \citet{Ludwig:1999p7606}, we performed functional fits
of the mixing-length parameters and the mass mixing-length parameter
with the $\teff$ and $\logg$ for the different metallicities individually.
We transformed the stellar parameters with $x=(\teff-5777)/1000$
and $y=\logg-4.44$, and fitted the values with a least-squares minimization
method for the functional basis

\begin{eqnarray}
f\left(x,y\right) & = & a_{0}+\left(a_{1}+\left(a_{3}+a_{5}x+a_{6}y\right)x+a_{4}y\right)x+a_{2}y.\label{eq:hgl}
\end{eqnarray}
The resulting coefficients, $a_{i}$, are listed in Table \ref{tab:hgl_fit}.
\begin{table*}
\caption{\label{tab:hgl_fit}Coefficients $a_{i}$ of the linear function $f$
(Eq. \ref{eq:hgl}) for $\amlt\left(\ssbot\right)$, $\amlt\left(\Delta s\right)$,
and $\am$ for different metallicities. In the last two rows, we list
the root-mean-square and maximal deviation of the fits.}

\begin{tabular}{ccccccccccc}
\hline\hline
Value & $\feh$ & $a_0$ & $a_1$ & $a_2$ & $a_3$ & $a_4$ & $a_5$ & $a_6$ & rms$\Delta$ & max$\Delta$ \\
\hline
 $\amlt(\ssbot)$ &       +0.5 &   1.973739 &  -0.134290 &   0.163201 &   0.032132 &   0.046759 &  -0.025605 &   0.052871 &      0.022 &      0.040 \\
 &       +0.0 &   1.976078 &  -0.110071 &   0.175605 &   0.003978 &   0.103336 &  -0.058691 &   0.080557 &      0.017 &      0.038 \\
 &       -0.5 &   1.956357 &  -0.133645 &   0.133825 &   0.027491 &   0.049125 &  -0.048045 &   0.057956 &      0.027 &      0.042 \\
 &       -1.0 &   1.969945 &  -0.143710 &   0.149004 &   0.001154 &   0.052837 &  -0.033471 &   0.037823 &      0.020 &      0.058 \\
 &       -2.0 &   2.010997 &   0.012308 &   0.160894 &  -0.041272 &   0.180486 &  -0.059577 &   0.074409 &      0.033 &      0.067 \\
 &       -3.0 &   2.133974 &   0.053307 &   0.222283 &  -0.192920 &   0.225412 &  -0.064937 &   0.027230 &      0.066 &      0.149 \\
\hline
 $\amlt(\dss)$ &       +0.5 &   2.060065 &  -0.075697 &   0.183750 &   0.018061 &   0.160931 &  -0.110880 &   0.164789 &      0.063 &      0.091 \\
 &       +0.0 &   2.077069 &  -0.079283 &   0.153376 &   0.041062 &   0.098795 &  -0.108972 &   0.137377 &      0.075 &      0.139 \\
 &       -0.5 &   2.080653 &  -0.117156 &   0.139250 &   0.105874 &   0.063015 &  -0.104596 &   0.143233 &      0.095 &      0.206 \\
 &       -1.0 &   2.131896 &  -0.135578 &   0.195694 &   0.039771 &   0.109232 &  -0.074565 &   0.110530 &      0.054 &      0.096 \\
 &       -2.0 &   2.229049 &  -0.068633 &   0.248141 &  -0.043729 &   0.229523 &  -0.088846 &   0.112805 &      0.056 &      0.136 \\
 &       -3.0 &   2.324527 &  -0.011662 &   0.293515 &  -0.171136 &   0.305021 &  -0.112595 &   0.077837 &      0.109 &      0.248 \\
\hline
 $\am$ &       +0.5 &   1.791089 &  -0.183788 &   0.179118 &  -0.022163 &   0.096536 &  -0.028233 &   0.054834 &      0.039 &      0.077 \\
 &       +0.0 &   1.832344 &  -0.177105 &   0.166634 &   0.011835 &  -0.002416 &  -0.030472 &   0.019225 &      0.023 &      0.045 \\
 &       -0.5 &   1.859980 &  -0.208802 &   0.154482 &   0.111923 &  -0.001357 &  -0.089213 &   0.105822 &      0.518 &      0.187 \\
 &       -1.0 &   1.897928 &  -0.208284 &   0.174666 &   0.035389 &   0.020293 &  -0.045907 &   0.031081 &      0.133 &      0.123 \\
 &       -2.0 &   1.959977 &  -0.255688 &   0.183739 &   0.032684 &   0.000570 &  -0.032134 &   0.000400 &      0.107 &      0.317 \\
\hline
\hline\end{tabular}
\end{table*}

\section{Addendum on MLT\label{app:Addendum_MLT}}

\subsection{mixing-length formulation\label{app:Mixing-length-formulation}}

In the framework of MLT, the convective flux is determined by 
\begin{eqnarray}
F_{\mathrm{conv}} & = & \left[\amlt c_{P}T\Delta/2\right]\rho v_{\mathrm{MLT}},\label{eq:convective flux}
\end{eqnarray}
with $c_{P}$ being the heat capacity, $\Delta$ the superadiabatic
energy excess, and $\amlt$ the adjustable\emph{ mixing-length parameter},
giving the mean free path of convective elements in units of pressure
scale height. The convective velocity is determined by 
\begin{eqnarray}
v_{\mathrm{MLT}} & = & \sqrt{\amlt^{2}gH_{P}\delta\Delta/\nu},\label{eq:convective velocity}
\end{eqnarray}
where $H_{P}$ is the pressure scale height, $\delta=-\left(\partial\ln\rho/\partial\ln T\right)_{p}$
the thermal expansion coefficient, and $\nu$ the energy dissipation
by turbulent viscosity. The superadiabatic excess is given by 
\begin{eqnarray}
\Delta & = & \frac{\Gamma}{\left(1+\Gamma\right)}\left(\nabla-\nabla_{\mathrm{ad}}\right),\label{eq:superadiabatic excess}
\end{eqnarray}
and the convective efficiency factor by 
\begin{eqnarray}
\Gamma & = & \frac{c_{P}}{8\sigma T^{3}}\tau_{e}\left(y+\tau_{e}^{-2}\right)\rho v_{\mathrm{MLT}},\label{eq:convective efficiency}
\end{eqnarray}
with the optical thickness $\tau_{e}$, and temperature distribution
$y$ of the convective element. The turbulent pressure 
\begin{eqnarray}
p_{\mathrm{turb}} & = & \beta\rho v_{\mathrm{turb}}^{2},\label{eq:turbulent_pressure}
\end{eqnarray}
can be included, but a depth-independent turbulent velocity, $v_{\mathrm{turb}}$
is assumed, which is the common approach for atmospheric modeling.
The resulting photospheric temperature stratifications are very similar
to the MARCS \citep{Gustafsson:2008p3814} and ATLAS models \citep{Kurucz:1979p4707,Castelli:2004p4949}.
In Paper I, we showed that below the surface, where convective energy
transport starts to dominate, the 1D models are systematically cooler
than the $\hav$ stratifications because of the fixed $\amlt$ with
$1.5$, in particular for hotter $\teff$.

\subsection{Influence of additional MLT parameters\label{app:Additional-MLT-parameters}}

\begin{figure}
\includegraphics[width=88mm]{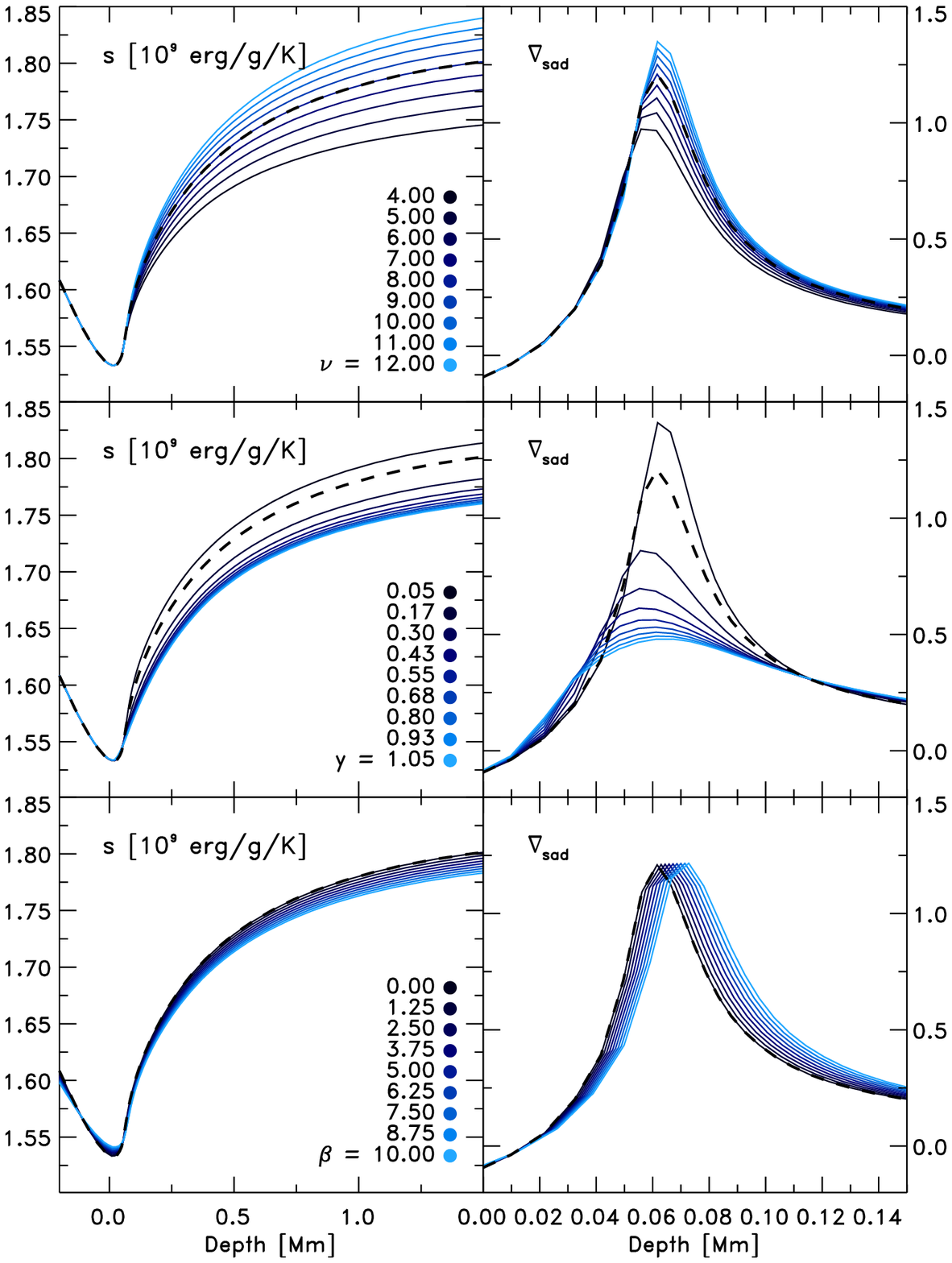}

\caption{Entropy and superadiabatic gradient vs.\ depth (left and right panel,
respectively) illustrating the influence of the additional MLT parameters
$\nu$, $y$, and $\beta$ (top, middle, and bottom panel, respectively),
the latter with the depth-independent $v_{\mathrm{turb}}=1\,\mathrm{km}/\mathrm{s}$.
The mixing-length is kept fixed at $\amlt=1.5$. We also included
the standard values of $\beta=0$, $\nu=8$ and $y=0.076$ (dashed
lines). Shown is the case for solar parameters.}

\label{fig:hidden_mlt} 
\end{figure}
In the formulation of \citet{Henyey:1965p15592} of MLT, there are
at least three additional free parameters apart from $\amlt$, which
usually are not mentioned explicitly, but are compensated for by the
value of $\amlt$. These are the scaling factor of the turbulent pressure,
$\beta$, the energy dissipation by turbulent viscosity, $\nu$, and
the temperature distribution of a convective element, $y$. The default
values are usually $\beta=1/2$, $\nu=8$ and $y=3/4\pi^{2}=0.076$
\citep[see][]{Gustafsson:2008p3814}. In many cases, the turbulent
pressure is neglected ($\beta=0$). In the notation of \citet{Ludwig:1999p7606},
these parameters would yield $f_{1}=\nu^{-1}$ and $f_{4}=y^{-1}$,
$f_{2}=1/2$ and $f_{3}=8/y$.

The turbulent pressure indirectly influences the $T$-stratification,
gradients, and hydrostatic equilibrium by reducing the gas pressure.
The parameter $\nu$ enters the convective velocity inverse proportionally,
$v_{\mathrm{MLT}}\propto\nu^{-1}$ (see Eq. \ref{eq:convective velocity}),
and since $v_{\mathrm{MLT}}\propto\amlt^{2}$, an increase in $\nu$
would have the same effect as a reduction in $\amlt$, i.e.\ $\nu\propto\ssbot$.
On the other hand, $y$ enters in the (nonlinear) convective efficiency
factor, $\Gamma$, for the superadiabatic excess (see Eq. \ref{eq:convective efficiency}),
and therefore $y$ is correlated with $\amlt$ in a more complex way.

Considering a variation of the three additional parameters in the
computation of the solar 1D model, we notice that the adiabatic entropy
value of the deep convection zone is altered significantly (see Fig.~\ref{fig:hidden_mlt}).
Furthermore, the two parameters $\nu$ and $y$ also change the entropy
jump and the superadiabatic temperature gradient, $\nsad$, and in
particular, its maximum of $\nsad$. The effect of the variation of
$y$ on the entropy stratification is similar to that by $\amlt$
(see Fig.~\ref{fig:match_alpha_mlt_atmo}). However, the entropy
of the deep convection zone exhibits a more nonlinear dependence with
the $y$ parameter. The increasing turbulent pressure with higher
$\beta$ changes the stratification only slightly, but shifts the
location of the maximum of $\nsad$ to the deeper interior. Towards
the optical surface the influence of the MLT parameters decreases,
as expected because of decreasing convective flux. A fine-tuning of
$\beta$, $\nu$ and $y$ is only useful when these parameters introduce
an independent influence on the mixing-length, since otherwise its
effects can be summarized in $\amlt$ alone.
\end{document}